\pdfoutput=1
\documentclass[multi]{cambridge7A}
\usepackage{natbib}
\usepackage{graphicx}
\begin{document}

\alphafootnotes

\author[R. I. Hynes]{Robert I. Hynes\footnotemark}

\chapter[Accretion in X-ray Binary Systems]{Multiwavelength
  Observations of Accretion in Low-Mass X-ray Binary Systems}

\footnotetext[1]{Louisiana State University,
 Department of Physics and Astronomy, 202 Nicholson Hall, Tower
 Drive, Baton Rouge, LA 70803, USA}

\arabicfootnotes

\begin{abstract}
  This work is intended to provide an introduction to multiwavelength 
  observations of low-mass X-ray binaries and the techniques used to
  analyze and interpret their data. The focus will primarily be on
  ultraviolet, optical, and infrared observations and their
  connections to other wavelengths.  The topics covered include:
  outbursts of soft X-ray transients, accretion disk spectral energy
  distributions, orbital lightcurves in luminous and quiescent states,
  super-orbital and sub-orbital variability, line spectra, system
  parameter determinations, and echo-mapping and other rapid correlated
  variability.
\end{abstract}

\section{Introduction}
\label{IntroSection}

The first X-ray binary to be observed and identified as such was
Scorpius X-1 \citep{Giacconi:1962a}, although several other systems
were known as optical stars or novae before this. Within a few years,
optical and radio counterparts to Sco X-1 were discovered
\citep{Sandage:1966a,Andrew:1968a}, and the topic has remained
multiwavelength in nature since then.

This work is intended to provide an introduction to some of the
observational characteristics of X-ray binaries suitable for a
graduate student or an advanced undergraduate. My aim was to produce a
primer for someone relatively new to the field rather than a
comprehensive review. Where appropriate I will also discuss techniques
for analysis and interpretation of the data. The focus is almost
exclusively on low-mass X-ray binaries, in which the accretion disk is
most accessible to multiwavelength observations, and is predominantly
biased towards ultraviolet, optical, and infrared observations, and
their relation to observations at other wavelengths. For a textbook
treatment of accretion astrophysics in general, the reader is referred
to \citet{Frank:2002a} and for more comprehensive reviews of X-ray
binaries to \citet{Lewin:1995a} and \citet{Lewin:2006a}.

We begin in Section~\ref{GeometrySection} by presenting an overview of
the main classes of X-ray binaries and their accretion geometries, and
in Section~\ref{TransientSection} describe observations of transient
systems. We will then address different types of observations in turn.
In Section~\ref{SEDSection} we examine expected and observed spectral
energy distributions, in Section~\ref{LightCurveSection} we move on to
consider orbital lightcurves, together with variability on
super-orbital and sub-orbital (but still relatively long) timescales.
Section~\ref{SpectroscopySection} then examines spectroscopic
observations, with a focus on determination of binary system
parameters, and finally Section~\ref{RapidSection} looks at shorter
timescale variations and in particular correlations between the X-ray
and UV/optical/IR behavior. For reference, we include a glossary of
notable objects in Section~\ref{ObjectSection} and of acronyms used in
Section~\ref{AcronymSection}.

\section{Classification and Geometry of X-ray binaries}
\label{GeometrySection}

The family of X-ray binaries is diverse. A single hierarchical
classification system is insufficient and depending on the question to
be asked one may desire to classify objects by the nature of their
donor star and mode of accretion, by the nature of the compact object,
or by whether their X-ray activity is persistent or transient.
Additionally, a subset of these objects produce relativistic jets and
are classified as microquasars

\begin{figure}
\includegraphics[width=3in]{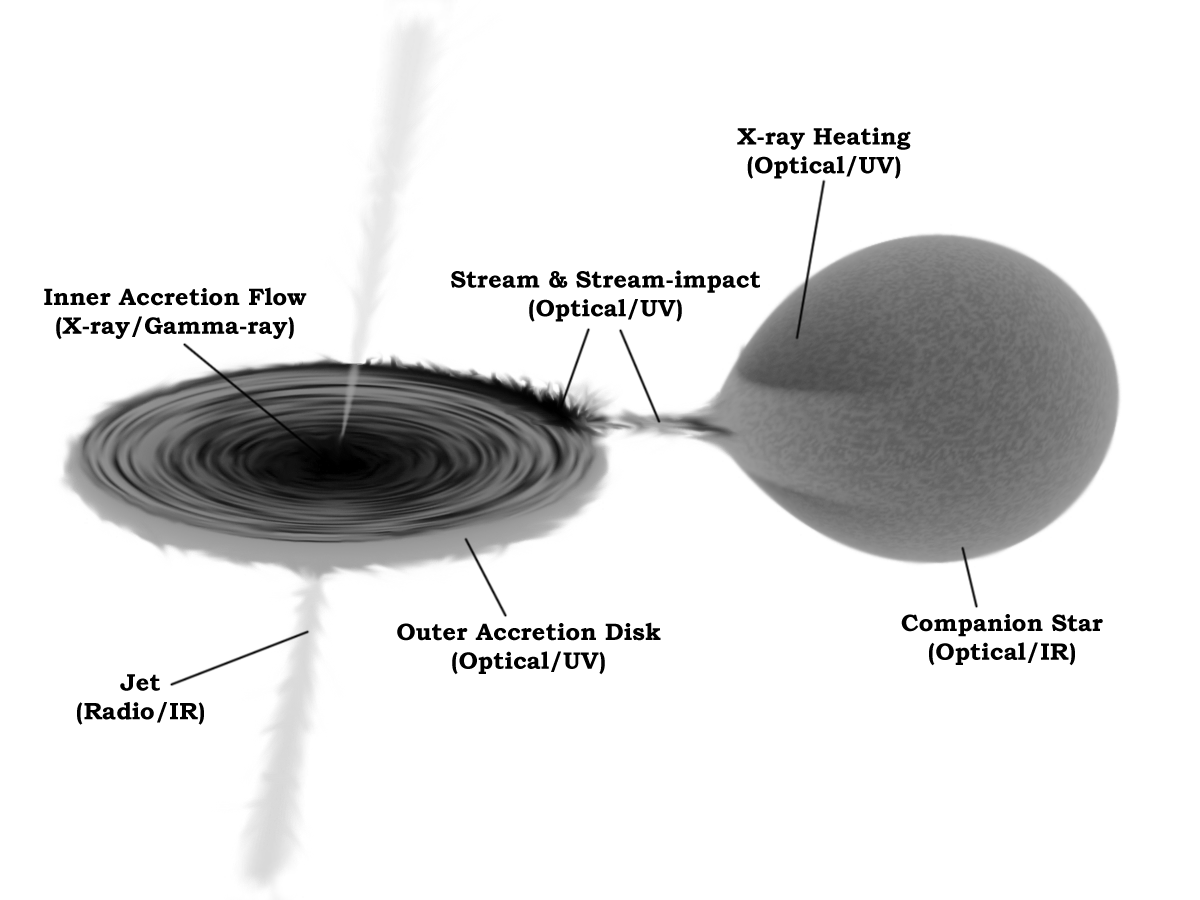}
\caption{Geometry of a Roche lobe overflow low-mass X-ray binary.  The
color scale is inverted with the brightest areas appearing darkest.}
\label{BinSimFig}
\end{figure}

The most common division made is into high-mass and low-mass X-ray
binaries: HMXBs and LMXBs. This is based crudely on the mass of the
donor star, with HMXBs typically having O or B type donors of mass
over 10\,M$_{\odot}$ and LMXBs having late-type G--M donors with
masses below $\sim1.5$\,M$_{\odot}$. Classification of those systems
with A or F donors has been vaguer, but these objects such as Her~X-1
and GRO~J1655--40 have often been grouped as intermediate-mass X-ray
binaries, or IMXBs. The donor mass also provides the primary division
in accretion mode, with LMXBs and IMXBs mostly accreting by Roche lobe
overflow (see Fig.~\ref{BinSimFig}) and HMXBs by either lobe overflow,
capture of wind from the donor star, or interaction of the compact
object with a circumstellar disk. A small group of LMXBs with low-mass
red giant donors do accrete from the donor wind. These are referred to
as symbiotic X-ray binaries \citep{Corbet:2008a}. Among the LMXBs,
those with exceptionally short periods and hydrogen deficient donors
are often broken out into the sub-class of ultracompact X-ray binaries
(UCXBs).

Another distinction can be made between systems containing black holes
and those containing neutron stars. For many purposes, this is more
important than the donor nature or accretion mode, and we observe many
similarities in X-ray behavior between black holes in LMXBs and HMXBs.

Observationally, an important difference is found between persistent
and transient sources. As time has passed this distinction has
blurred, and while there remain simple classical transients like
A\,0620--00 which undergo dramatic outbursts every few decades, other
systems seem to turn on into a semi-persistent state, or execute a
whole series of outbursts in succession after emerging from
quiescence. As our baseline extends, this kind of quasi-persistent, or
quasi-recurrent behavior with distinct on and off periods may become a
more common characteristic.

Finally, we should mention the category of microquasars. This term has
had varying usage, but in the context of X-ray binaries it usually
refers to those sources showing resolved, expanding, relativistic jets
such as GRS~1915+105 \citep{Mirabel:1994a}. Its usage has varied from
this fairly exclusive definition, to including all jet sources, to
potentially all black hole binaries, and possibly also some neutron
star sources.

\section{Transient X-ray binaries}
\label{TransientSection}

\subsection{Classical Soft X-ray Transients}

The outburst of the transient LMXB A\,0620--00 in 1975
(\citealt{Kuulkers:1998a} and references therein) opened up what was
to become a major area of X-ray binary research. It was not the first
transient X-ray source found, but was the first to be studied in great
detail, and the brightest yet seen.

Among the many reasons for the modern importance of transient systems,
their large dynamic range is invaluable. In a single object, on a
practical timescale of months to a few years we can watch the
evolution of the system through the full range of accretion states and
follow causal sequences between them. This is impossible with
persistent X-ray binaries and active galactic nuclei (AGN), both of
which individually sample a smaller range of parameter space leaving
us to attempt to build a complete picture from snapshots of different
objects. Another important characteristic of transient LMXBs is that
in quiescence their light usually becomes dominated by their companion
star. Radial velocity studies then allow measurement of system
parameters, and particularly the compact object mass. This is how we
know of the most compelling examples of stellar mass black holes, and
that the majority of known transient LMXBs are black hole systems. We
will consider system parameter determinations in more detail in
Section~\ref{ParameterSection}.

These transient LMXBs are often referred to as Soft X-ray Transients
(SXTs) based on the ultrasoft spectra that are sometimes seen in
outburst or Black Hole X-ray Transients (BHXRTs) due to their high
incidence of black holes. Among them is a subset which shows a
relatively orderly behavior with common features repeated between
several objects. The outbursts of these objects are known as FREDs --
Fast Rise Exponential Decay outbursts. FRED outbursts are often
considered to be typical of SXTs, but in fact there are many
exceptions. In the sample of outburst lightcurves compiled by
\citet{Chen:1997a}, there are both FREDs and irregular outbursts.
Since then, we have seen a preponderance of irregular outbursts and
the orderly FRED behavior now seems the exception rather than the
rule. Nonetheless, such a repeating pattern is a natural place to
begin in trying to understand their behavior and has provided the
benchmark for numerical simulations of outburst lightcurves (e.g.
\citealt{Cannizzo:1995a,Dubus:2001a,Truss:2002a}).

\begin{figure}[t]
\includegraphics[angle=180,width=3in]{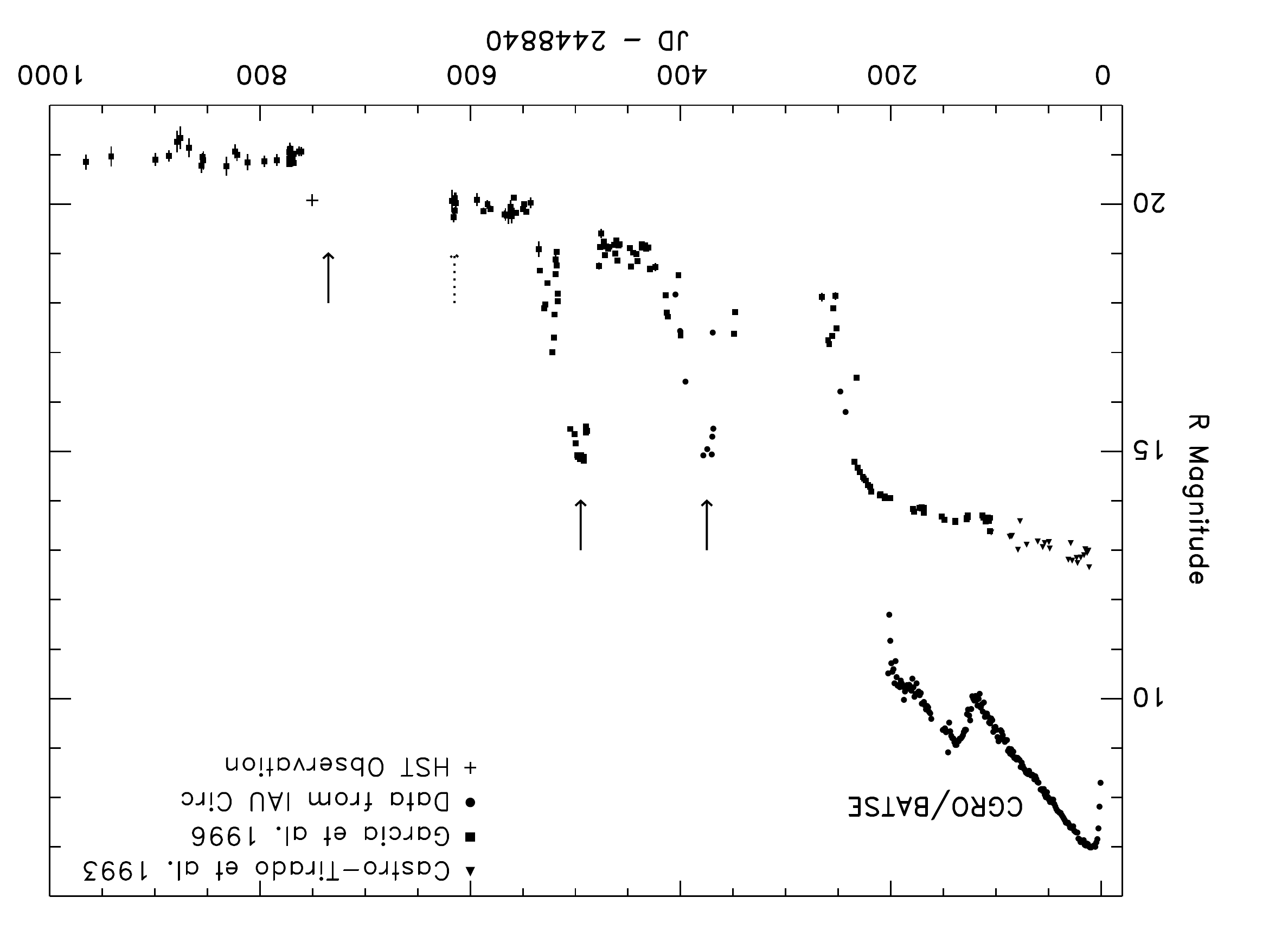}
\caption{Outburst lightcurves of GRO~J0422+32 adapted from
  \citet{Hynes:1999a}.  This shows many of the features of a canonical
SXT outburst; a fast rise, exponential decay, and mini-outbursts.
Note that the secondary maximum seen prominently in the CGRO/BATSE
lightcurve is not the one discussed in the text.  The one discussed
occurs earlier and is much less prominent in this object.}
\label{SXTFig}
\end{figure}

We show the hard X-ray (CGRO/BATSE) and optical lightcurves of
GRO~J0422+32 in Fig.~\ref{SXTFig} as an example of a FRED lightcurve
showing most of the common characteristics. The outburst exhibits a
very fast rise followed by a slow exponential decay for about 200
days. During the decay there are several secondary maxima. At the end
of the exponential decay comes a rapid drop-off. After this the
outburst was below the CGRO/BATSE threshold, but continued evolution
could be seen in the optical, with a continued decline interrupted by
at least two mini-outbursts.

The accepted framework for understanding SXT outbursts is believed to
be the disk instability model (DIM, \citealt{Lasota:2001a}) originally
developed to explain outbursts of dwarf novae (DNe). The latter
usually have a much shorter recurrence time and non-exponential
decays. The exponential decay in SXTs has been explained by
\citet{King:1998a} as a consequence of irradiation maintaining the
whole disk in the hot, high viscosity state of the DIM. The accretion
rate is proportional to the active disk mass, and so if the luminosity
traces the accretion rate, then the exponential decay in luminosity
reflects the decay of the disk mass. For short-period systems where it
is possible to maintain the whole disk in a high state in this way, it
is possible to accrete a large fraction of the disk mass during an
outburst. Consequently, the recurrence time is longer than in DNe,
where only a small fraction of the disk mass is accreted in a single
outburst.

Several explanations have been advanced for the secondary maxima, and
different features at different points in the decay may have different
explanations. In the earliest explanations, the turn-on of X-rays
during the outburst irradiated the companion star and stimulated a
burst of enhanced mass transfer into the disk. When this material
reached the compact object, the X-ray flux was further increased
\citep{Chen:1993a,Augusteijn:1993a}. \citet{King:1998a} instead
proposed that initially there is an outer region of the disk that
begins the outburst in the cool low viscosity state of the DIM.
Irradiation at the onset of the outburst then raises this to the high
state, and again, its effect on the X-ray lightcurve is delayed by one
viscous time. Finally \citet{Truss:2002a} modified this mechanism by
instead invoking the growth of tidal instabilities as a mechanism to
stimulate accretion from the outermost part of the disk.

The explanation for mini-outbursts remains less conclusive. This
phenomenon has been less commonly observed in SXTs, as it happens
later when the source has passed below the threshold of all-sky
monitors and so is typically mostly covered by sparser optical
coverage. These mini-outbursts are also characterized by a higher
ratio of optical to X-ray flux. The phenomenon closely resembles the
`echo-outbursts' seen in some cataclysmic variables (CVs) where on
occasion as many as six roughly uniformly spaced mini-outbursts have
been seen \citep{Patterson:1998a}. In attempting to explain
mini-outbursts, \citet{Hameury:2000a} again invoked irradiation,
\citet{Osaki:2001a} attributed them to failed viscosity decay attempts
at the end of the outburst, while \citet{Hellier:2001a} associates
them with tidal effects.

Once again, it should be emphasized that only a minority of SXT
outbursts follow the classic FRED form (and indeed that not all SXT
outbursts even exhibit the defining ultrasoft state;
\citealt{Brocksopp:2004a}). In many cases the irregular outbursts are
associated with longer period systems where it may not be possible to
raise all of the accretion disk into the hot state resulting in
interactions between the hot inner disk and the permanently cold outer
region. However, some short period systems such as XTE~J1118+480 also
exhibit irregular (and recurrent) outburst activity.

\subsection{Recurrent transients and semi-persistent systems}

It is likely all transient systems are recurrent, of course, but some
recur on shorter timescales than others. The best defined of these are
systems where outbursts occur once per orbital period, usually near
closest approach (periastron). The Be star plus neutron star HMXBs are
the most common example of this type, but Cir~X-1 is another eccentric
transient system.

Some transient LMXB sources emerge from quiescence to undergo multiple
recurrent outbursts like GRO~J1655--40, or turn on without
subsequently returning to quiescence like GRS~1915+105. GX~339--4 is
sometimes thought of as a rare persistent black hole but is actually
more like a very active recurrent transient, and given a longer
baseline may turn out to have been discovered in the midst of a
short-lived series of outbursts.

Among neutron stars, we see several that exhibit a similar kind of
semi-persistent behavior. 4U~2129+47 was identified as a 5.2\,hr LMXB
by \citet{Thorstensen:1979a} only to fade to quiescence in 1983
\citep{Pietsch:1983a}. EXO~0748--676 went into outburst in 1985
\citep{Parmar:1985a}, remained as an apparently persistent source for
24 years, before finally returning to quiescence in 2008.

This behavior among neutron stars provides a unique tool to probe the
properties of the neutron star. During an outburst, the crust is
heated by accretion and is no longer in thermal equilibrium with the
core. The cooling curve of the neutron star after outburst then
depends sensitively on the thermal conductivity of the crust, so these
observations can be used to test models for the structure and
composition of neutron star crust and the crust-core interface
\citep{Brown:2009a}.

\section{Spectral Energy Distributions}
\label{SEDSection}

\subsection{Overview}

X-ray binaries are multiwavelength objects with detectable emission
from radio to gamma-rays and at all wavelengths in between. The
spectral energy distribution (SED) measures the relative energy
contributed in different bands, and crudely characterizes the shape of
the spectra. The SED can inform us of the different components
emitting, in particular the disk, and the structure of those
components. Without additional spatial constraints, such as might be
provided by eclipse-mapping, we do not obtain any information about
the spatial arrangement of emitting regions, so there may be
degenerate models. In the simplest case of black body emission from an
accretion disk the information gained is really a measure of how much
emitting area is present at each temperature. With the common
assumptions that the disk is axisymmetric and that the temperature
increases monotonically inwards, however, we can translate this
information into the temperature as a function of radius. In practice,
it is more common to make these assumptions and fit simple derived
models to the SED, and so we will focus on the basis of these models.

At this point there are many different representations of the SED in
use and it is unlikely that a common standard will be adopted. It is
worthwhile to be familiar with the different conventions and to be
able to mentally transform between them. In order to show the order of
magnitude range of values typically present, almost all authors use
logarithmic axes for both the energy/wavelength axis and for flux. In
the X-ray binary community it is common to use an energy-like x-axis,
either based on photon frequency ($\nu$) or energy ($E$). The y-axis
is either the flux per unit frequency, $F_{\nu}$ (or equivalently per
unit energy, $F_E$), or this multiplied by the photon energy. Examples
of SEDs in the $\nu-F_{\nu}$ representation are shown in
Figs.~\ref{SEDFig} and \ref{SEDDataFig}. A straight $F_{\nu}$
representation makes it easier to estimate the spectral index,
$\alpha$, of a power law expressed as $F_{\nu}\propto \nu^{\alpha}$.
In particular, this representation is well suited to the typically
flat ($F_{\nu}\simeq$ constant) radio spectra often seen and
associated with jets. On the other hand, in a plot of $\log \nu
F_{\nu}$ vs. $\log \nu$, the height in $\nu F_{\nu}$ is a direct
measure of the amount of energy emerging per logarithmic frequency
interval at that frequency, hence the peak of an SED in $\nu F_{\nu}$
is the energetically most important component. A $\nu-\nu F_{\nu}$ SED
is also shown in Fig.~\ref{SEDFig}. The most common SED forms
encountered in X-ray binary work are then $\nu-F_{\nu}$, $\nu-\nu
F_{\nu}$, $E-F_E$, and $E-EF_E$, with the latter two most common in
dealing with X-ray and gamma-ray data. Wavelength based forms are
sometimes used, $\lambda-F_{\lambda}$ or $\lambda-\lambda
F_{\lambda}$. To mentally transform between them, it is helpful to
remember that a power-law of $F_{\nu}\propto \nu^{\alpha}$ becomes
$F_{\lambda}\propto \lambda^{2-\alpha}$ in this representation. The
final variation on these representations is unique to X-ray and
gamma-ray astronomy where the spectrum is sometimes described in an
$N_E-E$ representation where $N_E$ is the number of photons per unit
energy interval. Power-law spectra are then specified by the photon
index, $\Gamma$, where $N_E \propto E^{-\Gamma}$. Note the negative
sign in the definition. The photon index is related to the
$F_{\nu}-\nu$ spectral index by $\Gamma = 1-\alpha$.

\subsection{The Black Body Disk Model}

The spectral model that has most commonly been fitted to optical and
UV SEDs of X-ray binaries is a black body based model
\citep{LyndenBell:1969a,Shakura:1973a,Frank:2002a}. The disk is
assumed to be axisymmetric, with temperature increasing inwards.
Having defined the functional form of $T(R)$, the SED is evaluated by
summing emission from each concentric annulus. Simple limiting cases
can be derived analytically, but with realistic assumptions (for
example that both viscous and irradiative heating are present), a
numerical integration is more useful.

The problem can be defined for a power-law temperature distribution:
\begin{equation}
T(R)=T_0\left(\frac{R}{R_0}\right)^{-n}
\end{equation}
In the commonly discussed steady-state disk, $n=3/4$
\citep{Shakura:1973a}. We assume a local black body spectrum, $B_{\nu}$:
\begin{equation}
B_{\nu}(R) = \frac{2 h \nu^3}{c^2\left(e^{h\nu/kT(R)}-1\right)}
\end{equation}
and write the integrated SED of the disk as:
\begin{equation}
F_{\nu} = \frac{2\pi \cos i}{D^2}\int^{R_{\rm out}}_{R_{\rm in}} B_{\nu}(R)RdR
\end{equation}
where $D$ is the source distance and $i$ is the binary inclination.
This leads to the general solution:
\begin{equation}
F_{\nu}=\nu^{3-2/n}\frac{4\pi h^{1-2/n} \cos i k^{2/n}T_0^{2/n}R_0^2}{nc^2D^2}\int_{x_{\rm in}}^{x_{\rm out}}\frac{x^{2/n-1}dx}{e^x-1}
\end{equation}
where the substitution $x=h\nu/kT$ has been made. In the limiting case
of an unbounded disk extending from zero radius to infinity, the
integral is just a numerical value and the frequency dependence
extracted is $F_{\nu}\propto\nu^{3-2/n}$. For the Shakura-Sunyaev
disk, $n=3/4$, we obtain the well known result $F_{\nu}\propto
\nu^{1/3}$. If the disk is irradiatively heated then the simplest
assumption is that the irradiating flux drops off as the inverse
square of the radius leading to $T^4\propto R^{-2}$ and $T\propto
R^{-0.5}$. The resulting SED should then exhibit
$F_{\nu}\propto\nu^{-1}$, a red spectrum actually decreasing into the
UV.

\begin{figure}[t]
\includegraphics[angle=180,width=2in]{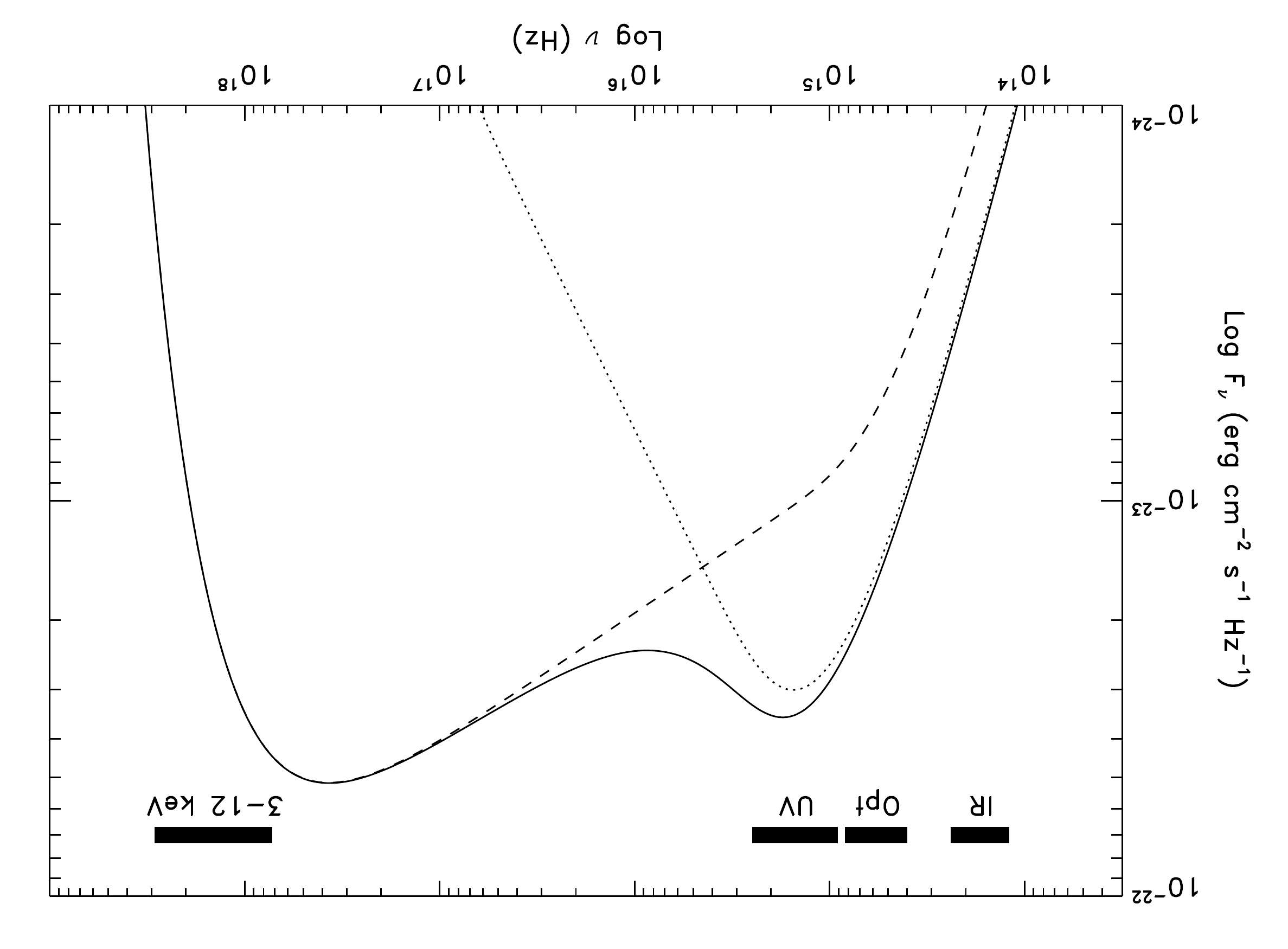}
\includegraphics[angle=180,width=2in]{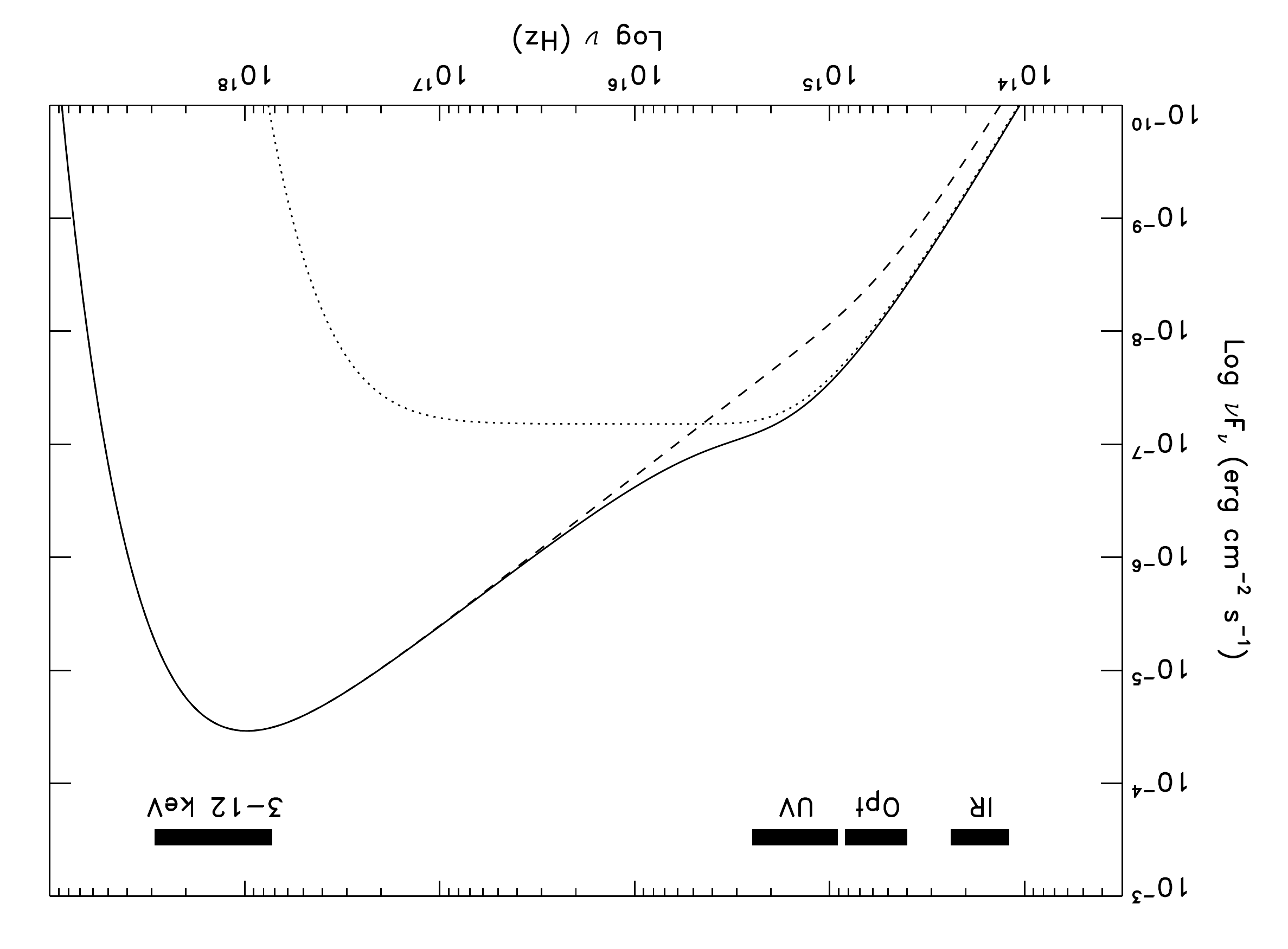}
\caption{Black body SED for cases with viscous heating (dotted),
  irradiative heating (dashed) and a combination of the two (solid).
  The left hand panel shows the $\nu-F_{\nu}$ representation, the
  right hand panel shows the same models in the $\nu-\nu F_{nu}$
  representation.}
\label{SEDFig}
\end{figure}

In practice, since the irradiation temperature $T_{\rm irr} \propto
R^{-0.5}$ drops off more slowly as a function of radius than the
temperature due to viscous heating, $T_{\rm visc}\propto R^{-3/4}$, we
expect the inner disk to be dominated by viscous heating and the outer
by irradiation, leading to a spectrum that turns over between the two
limits at some intermediate frequency (Fig.~\ref{SEDFig}).
Furthermore, the disk does not extend from zero radius to infinity,
and both inner and outer bounds are important in defining the
spectrum. The inner radius influences the X-ray spectrum and the outer
radius the UV and optical. Modeling of real X-ray data is more complex
than described here. Firstly, the temperature distribution of the
inner disk is modified by the boundary condition applied at the inner
edge of the disk. Secondly, Comptonization changes the X-ray spectrum
substantially (see Chapter by Done).

\subsection{Multiwavelength Observations of Black Holes in Outburst}

The first really high quality data set with which to test the picture
described above came from IUE and HST observations of the black hole
system X-ray Nova Muscae 1991 \citep{Cheng:1992a}. Multiple epochs of
data as the source faded revealed an approximately
$F_{\nu}\propto\nu^{1/3}$ spectrum consistent with a viscously heated
accretion disk. Taken at face value, however, this interpretation
requires extremely high accretion rates comparable to or in excess of
the Eddington limit (the accretion rate at which radiation pressure
balances gravity and can suppress further accretion).

\begin{figure}
\includegraphics[width=3in,angle=180]{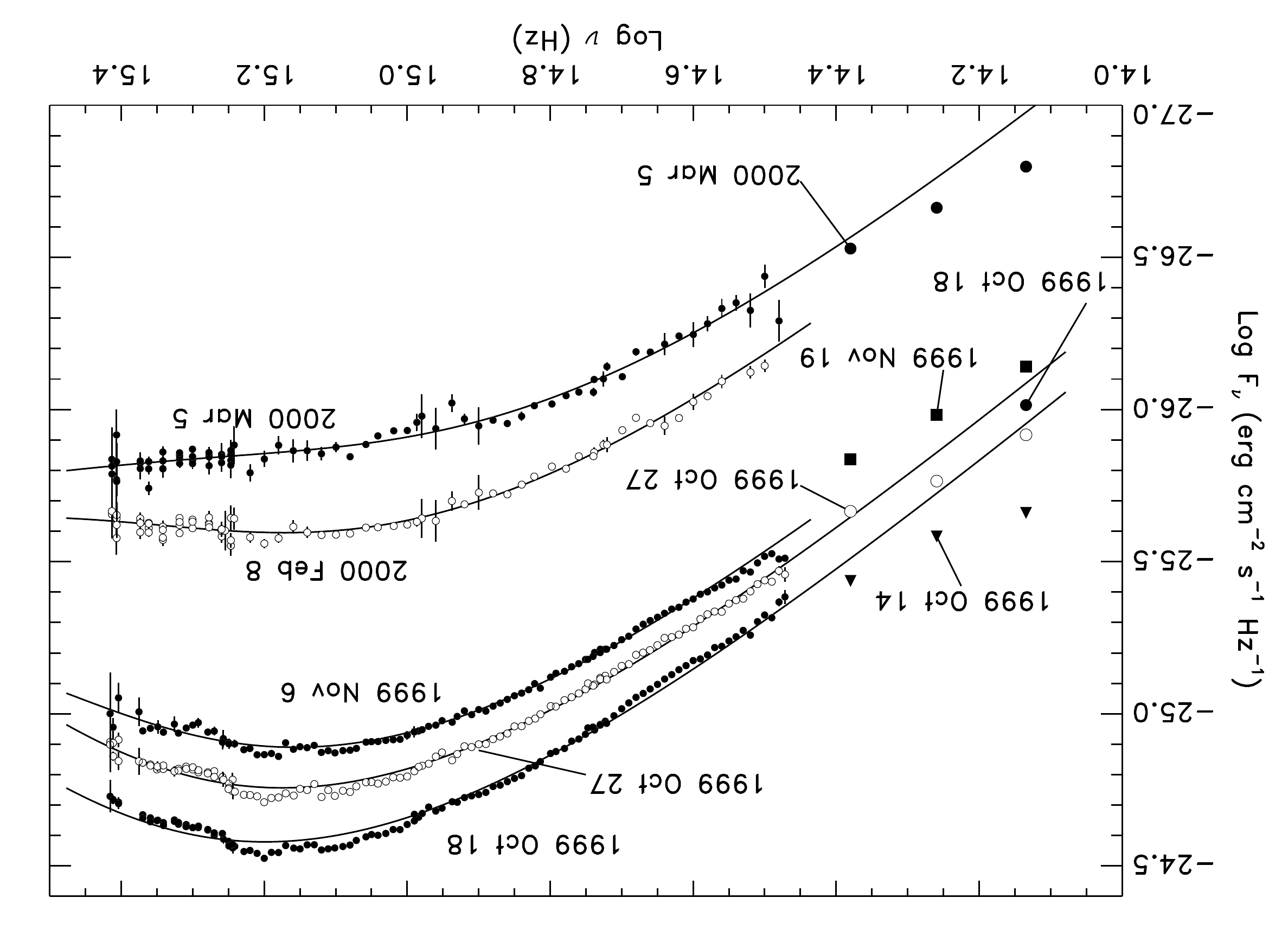}
\caption{Observed SEDs of XTE~J1859+226 from \citet{Hynes:2002a}.
  Different symbols are used to differentiate different observation dates.}
\label{SEDDataFig}
\end{figure}

Improvements in HST capabilities, and target of opportunity strategies
led to a much richer dataset on XTE~J1859+226 (Fig.~\ref{SEDDataFig};
\citealt{Hynes:2002a}), both in wavelength and temporal coverage.
Early observations showed a smooth spectrum with a broad hump in the
UV well fitted by black body disks with a relatively flat temperature
distribution consistent with an irradiated disk. Later in the
outburst, the far-UV spectrum transformed from a UV-soft form, where
$f_{\nu}$ declined with $\nu$ to a rising UV-hard form, consistent
with expectations of a purely viscously heated disk and similar to
that seen in X-ray Nova Mus 1991.

Other black hole transients have had less complete coverage but show
similar characteristics. A\,0620--00 showed a UV-hard SED resembling
X-ray Nova Mus 1991 or XTE~J1859+226 at late times, while GRO~J0422+32
was UV-soft like XTE~J1859+226 at early times (\citealt{Hynes:2005a}
and references therein).

It should be emphasized that the primary diagnostic between the
UV-hard and UV-soft states (and by inference $R^{-3}$ and $R^{-2}$
heating) is the turnover or its absence in the UV, and the slope of
the far-UV spectrum. \citet{Hynes:2005a} showed SEDs of these systems
and others based only on the optical portion of the SED, and no such
discrimination was possible. To obtain a reliable SED requires not
only good coverage in the satellite UV (for which HST is ideal, but
Swift/UVOT lacks far-UV capability), but also a good understanding of
the effect of interstellar reddening on the SED. For a discussion of
the latter, see \citet{Fitzpatrick:1999a} and references therein.

\subsection{Irradiation of Disks}

The amount of irradiation of the accretion disk that is actually
expected has been examined by \citet{Dubus:1999a} and expressed
somewhat differently by \citet{Dubus:2001a}. They express the local
irradiation temperature for a disk element at radius $R$ as:

\begin{equation}
\sigma T_{\rm Irr}^4 = \mathcal{C} \frac{L_{\rm X}}{4\pi R^2}
\end{equation}

where $\sigma$ is the Stefan-Boltzmann constant, $L_{\rm X}$ is the
irradiating (X-ray) luminosity, and $\mathcal{C}$ is a dimensionless
measure of the efficiency of irradiation. $\mathcal{C}$ parametrizes
our ignorance of the illumination geometry and the (energy-dependent)
albedo of the disk element. Empirically, \citet{Dubus:2001a} find that
a value of $\mathcal{C}\simeq5\times10^{-3}$ is consistent with
observations.

One difficulty with this simple picture that has been appreciated for
some time is that models of disk structure usually predict a disk
profile that is convex as a function of radius (\citealt{Dubus:1999a}
and references therein). Such a disk should be self-shielding and the
inner portion should be unable to irradiate the outer regions, in
spite of considerable evidence (such as echo-mapping;
Section~\ref{RapidSection}) that accretion disks do experience
irradiation. \citet{Dubus:1999a} suggest that either X-rays are
scattered by material out of the plane or that warping of the
accretion disk (see Section~\ref{WarpSection}) can expose portions of
the outer disk to irradiation. In this case, $\mathcal{C}$ also
parametrizes our ignorance about how irradiation of the disk is
mediated. The caveat here is that for many accretion geometries,
$\mathcal{C}$ itself becomes a function of radius.

\begin{figure}
\includegraphics[width=2in]{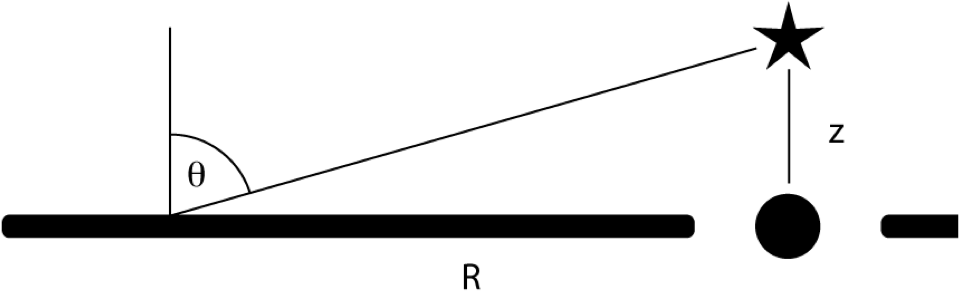}
\caption{Irradiation geometry appropriate for a `lamp post' above the
  accretion disk.}
\label{LampPostFig}
\end{figure}

One example where the radial dependence of irradiation can be
significantly modified is important to discuss, and to highlight how
an inferred temperature distribution may not be sufficient to
constrain the underlying astrophysics. This is the case of irradiation
by a `lamp post' above the disk, for example flares in a vertically
extended corona, or back-irradiation from a jet. This case is well
known in the AGN and young star communities where it is sometimes
considered the normal irradiation geometry. It is illustrated in
Fig.~\ref{LampPostFig}. In this case, the intensity of irradiating
flux decreases according to the inverse square law as considered
above, but in addition the angle of incidence becomes steeper at
larger radii resulting in spreading the irradiation over a larger
area. This introduces another factor proportional to $\cos
\theta\simeq z/R$ for $z \ll R$ (where $z$ is the height of the
irradiating source above the disk). The additional $1/R$ dependence
results in heating varying as $R^{-3}$ just as for the viscously
heated steady state accretion disk. It is therefore possible that the
UV-hard spectra discussed above could also result from irradiation,
and this may avoid the difficulty noted, for example in the case of
X-ray Nova Muscae 1991, where purely viscous heating would require
accretion at or above the Eddington limit.

\subsection{Evidence for Jets in Spectral Energy Distributions}
\label{JetSection}

\begin{figure}
\includegraphics[width=3in]{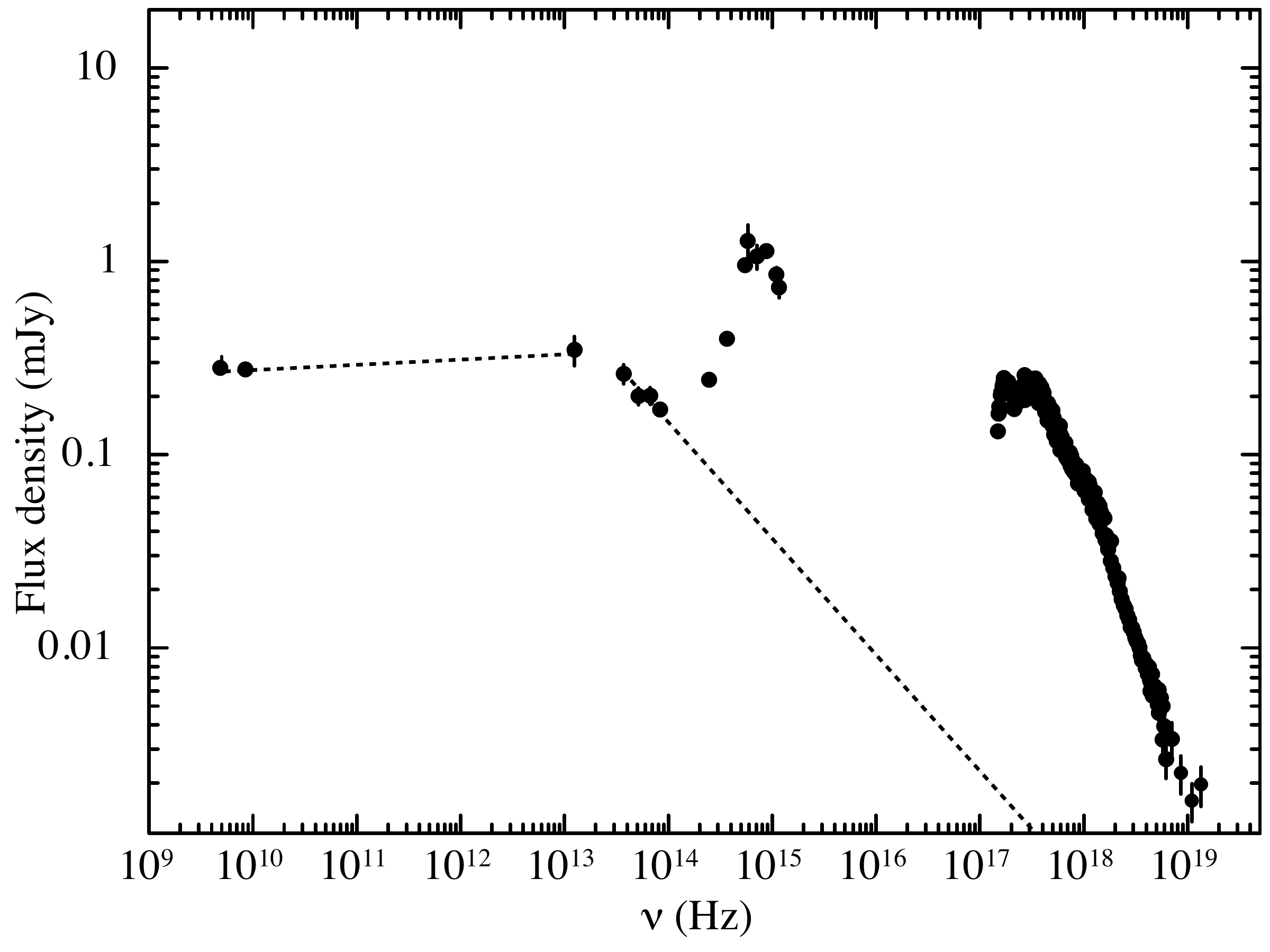}
\caption{SED of 4U~0614+091 adapted from \citet{Migliari:2010a}.}
\label{MigliariFig}
\end{figure}

Not all emission in the spectral energy distributions of LMXBs
originates in the disk. There is now very persuasive evidence in a
number of black hole LMXBs for IR synchrotron emission from a jet.
This is seen as either a flat IR spectrum extending to lower
frequencies than possible for the disk (XTE ~J1118+48;
\citealt{Hynes:2000a}) or even a two component spectrum with a red IR
component from the jet and a blue disk component (GX~339--4;
\citealt{Corbel:2002a}). In both cases the IR flux is comparable to
that seen in the radio as expected from extrapolation of the
characteristic flat spectrum of a compact radio jet
\citep{Blandford:1979a}. More recently a similar SED has also been
seen in the neutron star LMXB 4U~0614+091 (Fig.~\ref{MigliariFig};
\citealt{Migliari:2006a,Migliari:2010a}). The second observation was
based on quasi-simultaneous observations of a persistent source and so
is a robust measure of the SED.

Another signature expected for synchrotron emission from a jet would
be polarization. To date no large IR polarizations have been found,
although \citet{Shahbaz:2008a} saw polarizations in Sco~X-1 and
Cyg~X-2 of a few percent which increased with wavelength as would be
expected from the increasing fractional contribution of a jet at
longer wavelengths, and which were not consistent with expected
interstellar polarization.

\subsection{Quiescent SEDs}
\label{QuiescentSEDSection}

Observations of quiescent LMXBs also reveal emission across most of
the observable range, although radio detections are extremely
challenging and those at gamma-ray energies have been impossible so
far. A major complication in obtaining SEDs of the accretion emission
in quiescent LMXBs is that the companion star can completely dominate
the optical and IR range of the spectrum. Unfortunately, the brightest
quiescent system, V404~Cyg, also exhibits almost complete dominance of
the UV to near-IR SED by the companion \citep{Hynes:2009a}.

The accretion flow is clearly detected at X-ray energies in many
systems, and appears to be characterized by a soft power-law of photon
index $\Gamma\sim2$, corresponding to $F_{\nu} \propto \nu^{-1}$
(e.g.\ \citealt{Kong:2002a}). This almost certainly originates from
the inner region of the accretion flow which is believed to form a
hot, evaporated, advective flow. Spectral models of such flows
variously attribute the X-ray emission to either bremmstrahlung or
Comptonization, but there remain many uncertainties in these models
and there is no uniquely accepted solution
\citep{Narayan:1996a,Narayan:1997a,Esin:1997a,Quataert:1999a}. X-ray
emission lines could provide a valuable discriminant
\citep{Narayan:1999a}, but the most sensitive observations of a
quiescent SXT, XMM-Newton observations of V404~Cyg, yielded only upper
limits on the iron K$\alpha$ emission line that is expected to be
strongest \citep{Bradley:2007a}.

A UV excess is sometimes detectable
(\citealt{McClintock:1995a,McClintock:2000a,McClintock:2003a}; Hynes
et al.\ in preparation). The spectral shape can be characterized as a
hot quasi-blackbody of temperature $\sim10,000$\,K. Our best
explanation for this component currently is that it either originates
from the accretion stream-impact point, or from a relatively hot
region of the disk (hot enough that it should be locally in the high
state of the DIM).

A mid-IR excess is also seen \citep{Muno:2006a} but the origin remains
debated. In some sources such as V404~Cyg, the excess is rather
subtle, but in others like A\,0620--00 and XTE~J1118+480 it is very
pronounced and clearly real. The favored explanation of
\citet{Muno:2006a} for most sources was that it originates in a {\em
  circumbinary} accretion disk comprising material lost during
outbursts or even the original supernova that formed the compact
object. At longer wavelengths, radio emission appears to exhibit a
flat spectrum as is often seen in more luminous states
\citep{Gallo:2005a}. The most natural explanation for this is that a
weak jet continues in quiescence. It is even possible that this
dominates the quiescent energy budget with much of the accretion power
carried away in the bulk kinetic energy of the jet
\citep{Fender:2003a}. Interestingly, an extrapolation of the flat
radio spectrum into the mid-IR region comes reasonably close to the IR
excess flux, leading \citet{Gallo:2007a} to interpret the mid-IR
excess as due to the jet instead of a circumbinary disk, resulting in
a situation similar to the hard state systems discussed in
Section~\ref{JetSection}.

\section{Light Curves}
\label{LightCurveSection}

\subsection{Ellipsoidal Variations}
\label{EllipsoidalSection}

\begin{figure}
\includegraphics[width=2in,angle=180]{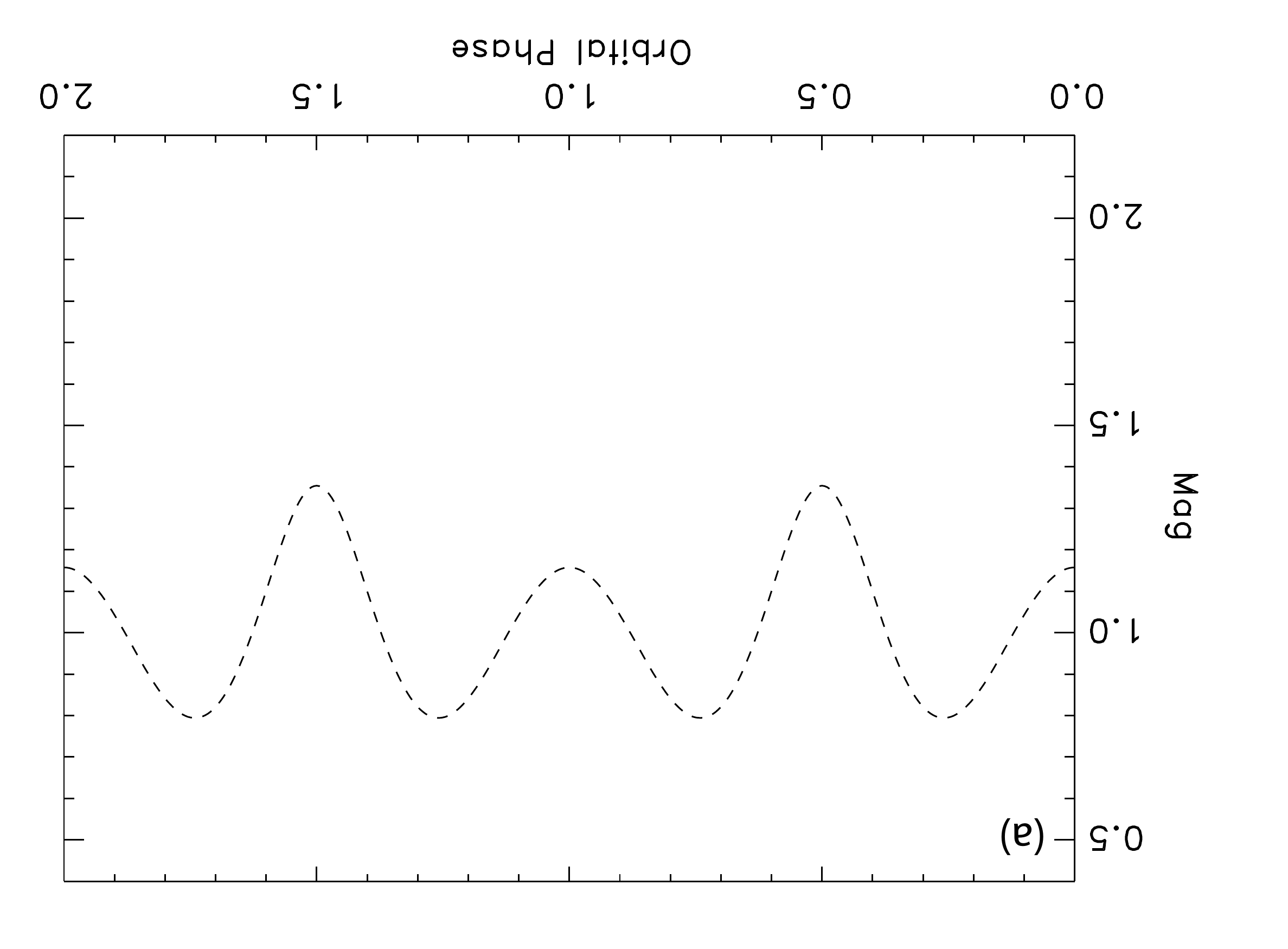}
\includegraphics[width=2in,angle=180]{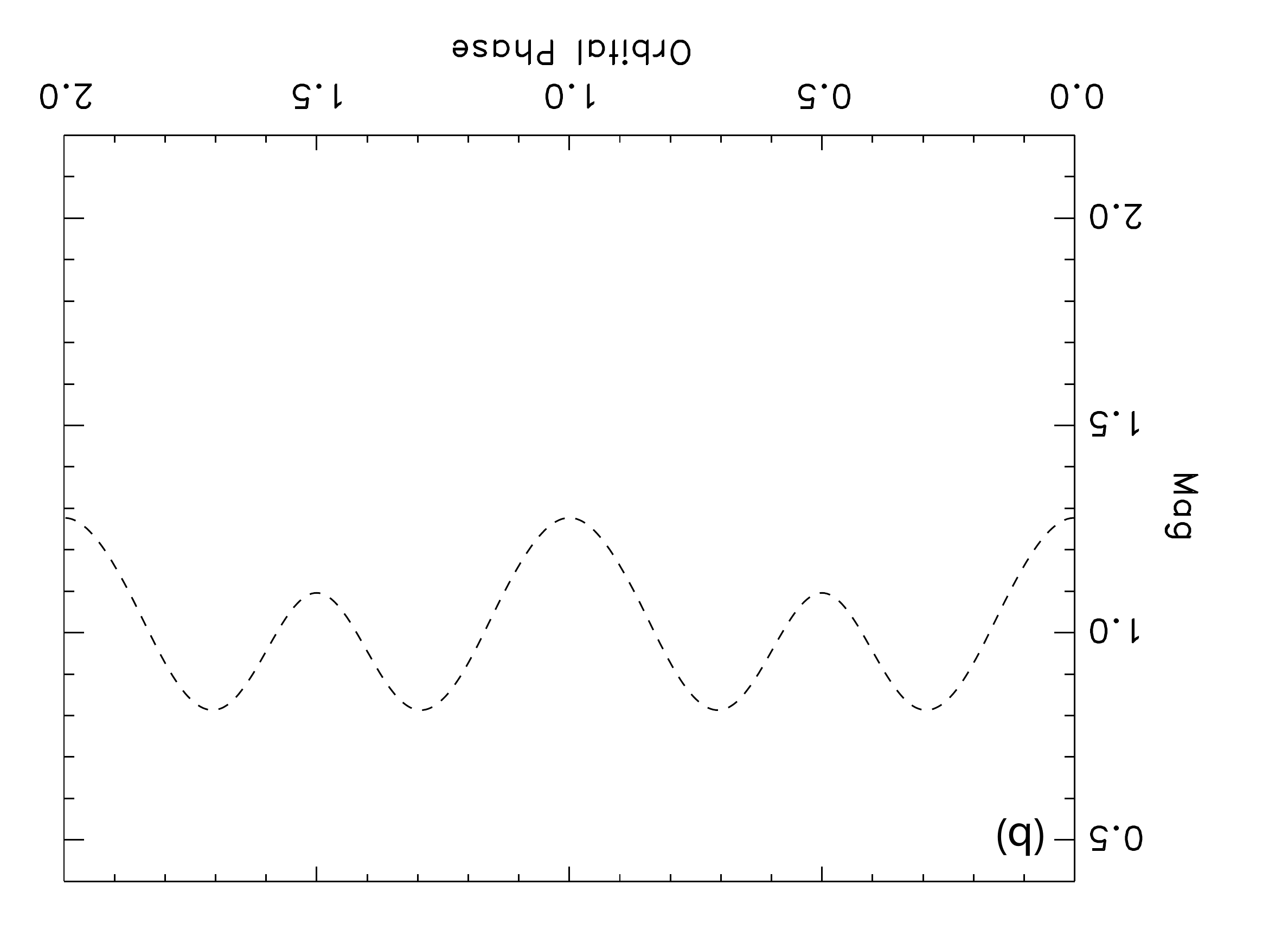}
\includegraphics[width=2in,angle=180]{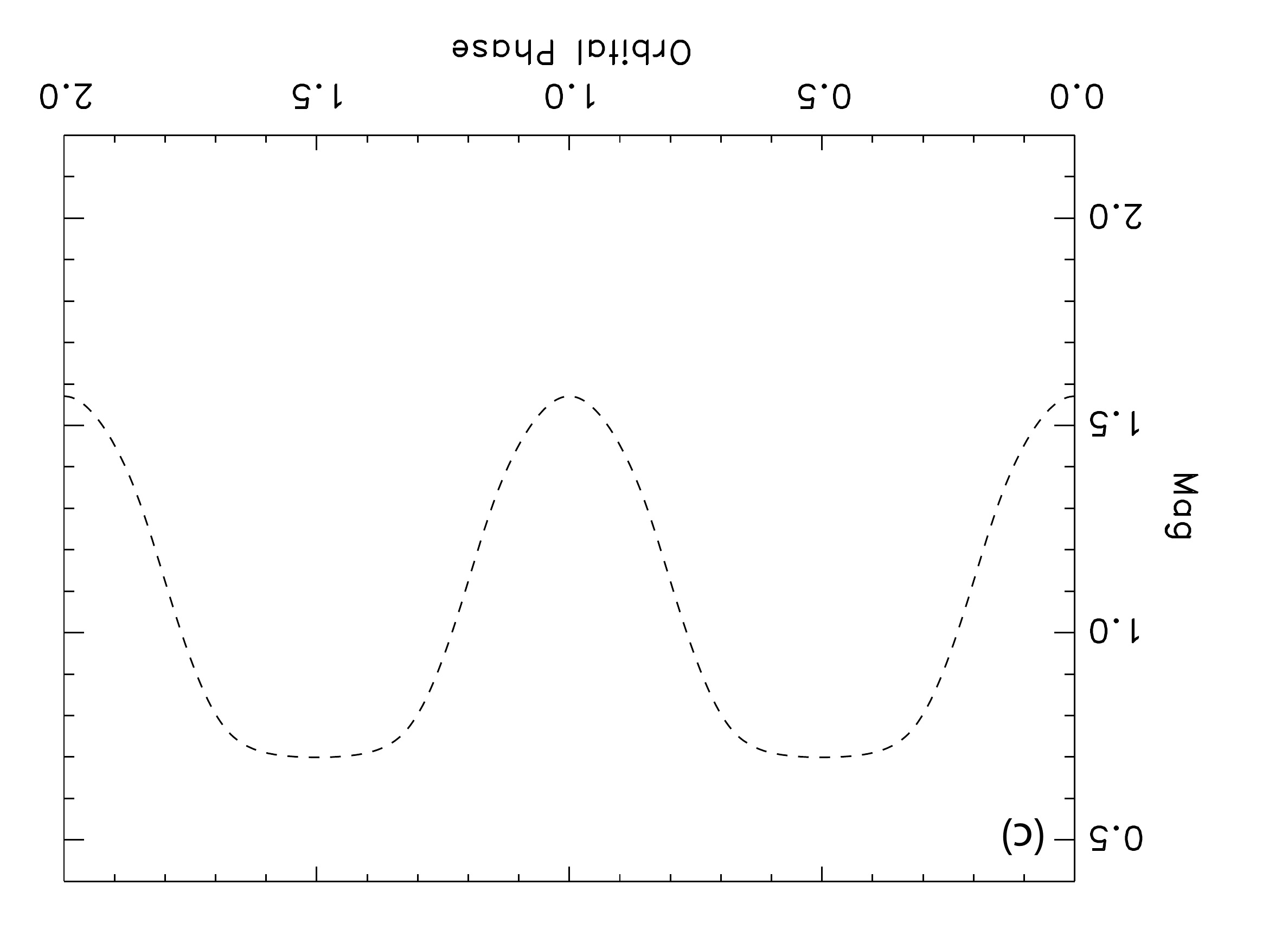}
\includegraphics[width=2in,angle=180]{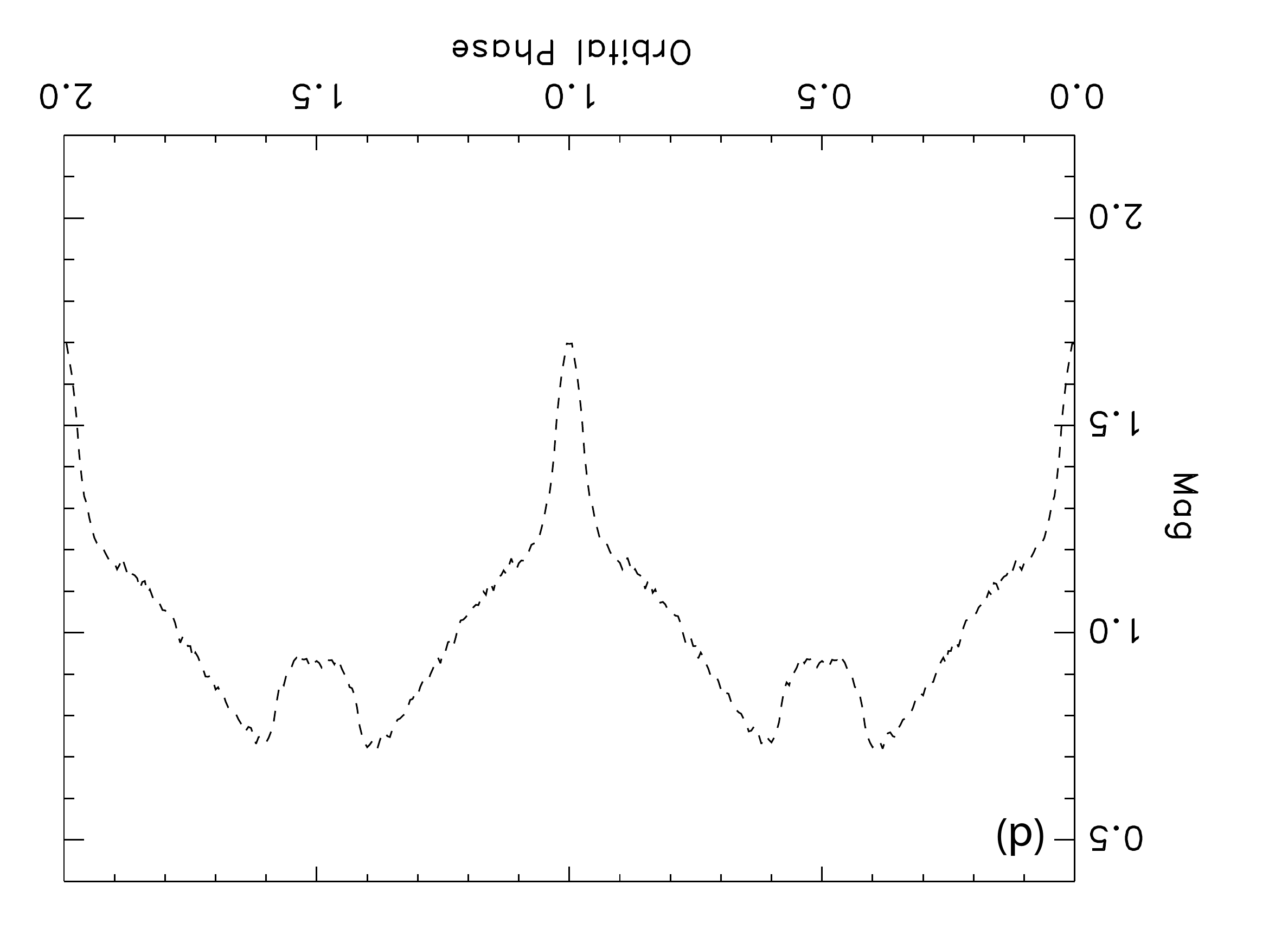}
\caption{Orbital lightcurves of LMXBs. a) Pure ellipsoidal
 variations. b) Ellipsoidal variations plus weak irradiation of the
 donor.  c) Strong irradiation of the donor.  d) Irradiation of disk
 and donor with mutual eclipses.  These figures were generated by the 
XRbinary code written by E. L. Robinson.}
\label{LightCurveFig}
\end{figure}

The simplest form of optical lightcurve of an LMXB is that from
ellipsoidal variations in quiescence where the light is often
dominated by the companion star. This star is tidally distorted into a
tear-drop shape. When viewed side-on at orbital phases 0.25 and 0.75,
we see a large cross-sectional area and hence maximum light. When
viewed end-on at phases 0.0 and 0.5 we see a smaller cross-sectional
area and hence light. In essence, ellipsoidal variations then take a
near-sinusoidal form with two cycles per binary orbit. An additional
complication is that the surface of the companion is not uniformly
bright due to gravity darkening \citep{vonZeipel:1924a}. Less flux
emerges from regions of the companion with a lower surface gravity,
and so these regions have a lower effective temperature. In the
absence of significant X-ray heating of the donor star, this results
in maximum surface temperature at the poles and minimum at the inner
Lagrangian point facing the compact object. This breaks the symmetry
between phase 0.0 and 0.5 (but not between 0.25 and 0.75).
Consequently ellipsoidal variations should exhibit two equal maxima
but unequal minima, with the phase 0.5 minimum being deeper. A model
ellipsoidal lightcurve is shown in Fig.~\ref{LightCurveFig}a.

\begin{figure}
\includegraphics[width=2in]{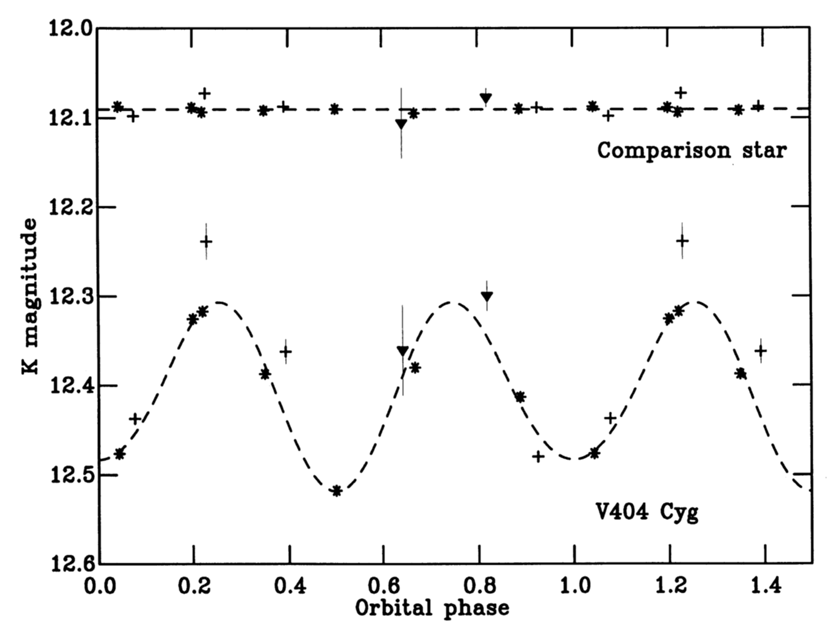}
\includegraphics[width=2in]{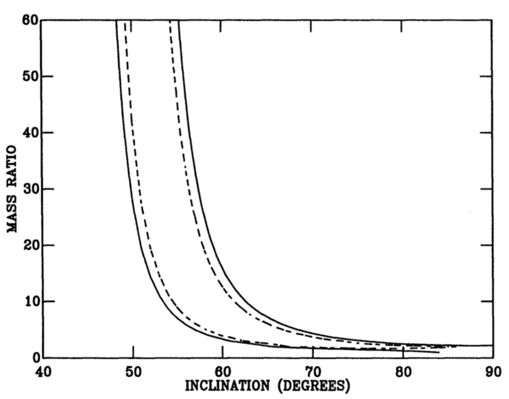}
\caption{Left: IR ellipsoidal modulations in the quiescent SXT
  V404~Cyg with a fitted model.  Right:  Derived joint constraints on
  inclination and mass ratio.  Both figures are from \citet{Shahbaz:1994a}.}
\label{ShahbazFig}
\end{figure}

The amplitude of ellipsoidal modulations is mostly determined by the
binary inclination, with a lesser dependence on mass ratio and an even
weaker dependence on stellar temperature and surface gravity (via limb
darkening changes). This has made measurement of ellipsoidal
modulations the technique of choice for determining inclinations of
non-eclipsing quiescent LMXBs. An example is shown in
Fig.~\ref{ShahbazFig}.

There are several caveats to this approach. The first and most obvious
is that the observed variations must actually be produced by
ellipsoidal modulations. Other mechanisms for periodic variability
include superhumps (see Section~\ref{SuperhumpSection}), variations in
the visibility of the accretion stream impact point, and starspots on
the companion star. There are certainly lightcurves which cannot fully
be explained by ellipsoidal variations. For example many studies have
now been made of A\,0620--00 which find that the lightcurves clearly
change in morphology from epoch to epoch, often showing asymmetric
maxima
\citep{McClintock:1986a,Haswell:1993a,Gelino:2001a,Cantrell:2008a}.
The authors have variously invoked state changes, disk asymmetries,
star spots, and eclipses to explain the behavior seen. While the
system is no longer believed to be at a high enough inclination to
eclipse, it is likely that several of the other effects may all play a
role.

Even in the absence of periodic contamination of the ellipsoidal
lightcurves, the presence of non-variable disk emission will dilute
the ellipsoidal variations. This will make the amplitude appear
smaller leading to an underestimate of the binary inclination and
ultimately an overestimate of the compact object mass. It is often
argued that this contamination can be minimized by measuring IR
ellipsoidal variations (e.g.\ \citealt{Gelino:2001a}), but an IR
contamination from a cool disk might well be {\em expected} in
quiescence \citep{Hynes:2005b} and the presence of IR excesses in
quiescent SXTs discussed in Section~\ref{QuiescentSEDSection}
reinforces this expectation. To attempt to compensate for this, one
can try to measure the amount of disk light by measuring dilution of
the photospheric absorption lines of the companion star; this is
termed the veiling. While some measurements suggest small or
undetectable IR veiling, e.g.\ V404~Cyg \citep{Shahbaz:1996a} in other
objects the IR veiling can be significant \citep{Froning:2007a}. In
any case, we know that flickering light is present in quiescence which
can have large amplitudes (e.g.\ \citealt{Hynes:2003b}), that IR
flickering can be detectable \citep{Reynolds:2007a}, and that longer
timescale changes in the overall flux level also occur
\citep{Cantrell:2008a}. These facts suggest that even if we believe
the veiling can be measured reliably, it is strictly only valid if
measured simultaneously with the ellipsoidal variations, and ideally
during the passive state identified by \citep{Cantrell:2008a}.

\subsection{X-ray Heating Effects}

As we move from quiescence into luminous or outburst states, the
accretion rate onto the compact object, and hence the X-ray
luminosity, increases. The first effect on the lightcurve is that the
temperature of the inner face of the companion star is raised by
irradiation from the increased X-ray luminosity. This offsets the
gravity darkening effect and will first weaken the phase 0.5 minimum,
then reverse the depths of the minima (so that phase 0.0 is the deeper
minimum). With strong enough heating, such as for typical short period
luminous LMXBs, the ellipsoidal effects are no longer discernible and
we see a single-humped modulation with a {\em maximum} at phase 0.5
and a minimum at phase 0.0. This progression from pure ellipsoidal
modulations to strong heating of the companion star is illustrated in
Fig.~\ref{LightCurveFig}a--c.

This effect is commonly observable in neutron star LMXBs allowing the
orbital period to be determined from optical photometry. In many cases
in the absence of X-ray eclipses or dips, this may be the first or
only evidence for the orbital period. For example, in Sco~X-1 the
orbital period of 0.78\,days was first identified in this way
\citep{Gottlieb:1975a}.

\subsection{Eclipses}
\label{EclipseSection}

If an LMXB has a high enough inclination, then eclipses of the
accretion disk by the companion star can occur, sharpening the minimum
at phase 0.0, while eclipses of the irradiated face of the companion
star by the disk may introduce a secondary eclipse at phase 0.5
(Fig.~\ref{LightCurveFig}d). Real lightcurves can look very much like
this model, for example XTE~J2123--058 \citep{Zurita:2000a}. At higher
inclinations self-obscuration of the disk becomes an important factor
and additional structure appears in the lightcurves reflecting
deviations from axisymmetry of the disk. The most pronounced
manifestation of this is X-ray dipping discovered in 4U~1915--05
\citep{White:1982a} and subsequently seen in about ten other neutron
star LMXBs. X-ray dips are characterized by rather irregular dipping
usually strongest around phases 0.7--1.0 likely indicating absorption
by inhomogeneous material associated with the accretion stream impact
point and/or material from it that overflows the disk. At even higher
inclinations the central X-ray source is {\em permanently} obscured by
the disk structure resulting in an accretion disk corona (ADC) source
where only scattered X-rays are seen. The prototypical ADC source is
2A~1822--371 \citep{Mason:1980a}; see \citet{Bayless:2010a} for a
recent multiwavelength study. Such sources are characterized by a much
lower ratio of X-ray to optical brightness than lower inclination
objects, and by broad partial X-ray eclipses rather than narrow total
ones.

Eclipsing systems provide much more precise constraints on the binary
inclination, and so have the potential to yield more accurate system
parameters than are possible with ellipsoidal variations alone.
Unfortunately, no eclipsing black hole LMXBs are known in our Galaxy
so this potential benefit for black hole mass determination has yet to
be fully realized. The only eclipsing black hole candidate known with
confidence is the HMXB, M33~X-8 \citep{Orosz:2007a}. The lack of
eclipsing black hole systems is unlikely to be coincidence, and
reflects a selection effect leading to an absence of black hole LMXBs
with inclination $i>75^{\circ}$. It is proposed that higher
inclination black hole systems are ADC sources where much of the X-ray
luminosity is hidden from view and so we are less likely to detect
them \citep{Narayan:2005a}. Neutron star ADC sources have been found
in spite of this selection effect because they are persistently active
whereas black holes are usually transient.

In CVs, eclipsing systems facilitate eclipse-mapping of the accretion
disk emission structure and radial temperature dependence (see
Warner's review in Chapter by Warner). This tomographic
eclipse-mapping methodology developed for CVs depends on the known
geometry of the companion Roche lobe occulting the disk. It has not
proved directly applicable in LMXBs because of the disk
self-obscuration effects together with the additional light from
heating of the companion star. Instead, LMXB lightcurves have been
interpreted by developing a model of the binary geometry, including
irradiation of the disk and companion and possibly asymmetric
structure in the accretion disk and then predicting multiwavelength
lightcurves. The nature of the self-obscuration remains a matter of
debate. Early models interpreted it as an elevated rim to the
accretion disk. The major flaw of this explanation is that to maintain
outer disk material at this elevation would require temperatures much
higher than expected, so such material could not be in hydrostatic
equilibrium. \citet{Bayless:2010a} instead suggest that the obscuring
material is a wind from the inner disk, and so is not in hydrostatic
equilibrium.

Another benefit to eclipsing LMXBs is that eclipses can provide very
precise timing of the orbital period, with ingress and egress times of
the total eclipse of the neutron star lasting just a few seconds. The
duration here reflects the scale height of the companion star
atmosphere, which is larger than the neutron star size. The best
studied case by far is EXO~0748--676 \citep{Wolff:2009a}. In this
system, an ongoing eclipse monitoring program through the history of
RXTE, coupled with less well sampled observations from the preceding
decade, has resulted in an orbital period history with exquisite
precision. Surprisingly, rather than showing a gradual period
evolution, this monitoring has revealed distinct epochs of duration
5--10 years with no obvious changes in the X-ray behavior
corresponding to the abrupt transitions between epochs.
\citet{Wolff:2009a} attributed these changes to magnetic cycles in the
companion star which modify its internal structure, and hence
redistribute a small amount of angular momentum. In support of this
interpretation, similar behavior is seen in magnetically active RS~CVn
systems.

\subsection{Superhumps}
\label{SuperhumpSection}

It should not be assumed that all periodicities are actually orbital
modulations. The tidal interaction of the companion star with the
outer disk can excite eccentric modes within the disk. The most
commonly encountered mode is excited when the disk grows to a radius
where there is a 3:1 resonance between the local Keplerian frequency
and the orbital frequency \citep{Whitehurst:1991a}. This is possible
only in systems with relatively small mass ratios ($q=M_2/M_1<0.25$)
as larger mass ratios result in the disk being truncated inside the
3:1 resonance radius by tidal stresses from the companion star. In
even more extreme mass ratio systems, it may be possible to also
excite the 2:1 resonance. Once eccentricity develops, the eccentric
disk will slowly precess. When the eccentric disk reaches closest to
the companion star, it experiences increased tidal stresses and hence
more heating. This leads to variations in the optical and UV light
from the outer disk, termed superhumps, that occur at the beat period
between the orbital period and the precession period. Superhumps are
commonly seen in CVs, and are discussed in more detail by Warner in
Chapter by Warner, but they can occur in LMXBs too.

The most likely LMXBs to exhibit superhumps are black hole systems as
these tend to have more extreme mass ratios than most neutron star
LMXBs. \citet{ODonoghue:1996a} examined the cases for several black
hole transients and concluded that superhumps had definitely been seen
in two systems, GRO~J0422+32 and X-ray Nova Muscae 1991, and had
possibly also been seen in GS~2000+25. Unlike the CV case, however,
where superhumps are uniquely present during outburst with the orbital
period absent, in LMXBs both orbital and superhump modulations may be
present, at different times or even simultaneously. The superhump
period is very similar to the orbital period (to within a few percent)
and discrimination between the two can be difficult. More recent and
very interesting examples of superhumping LMXBs have included
XTE~J1118+480 \citep{Uemura:2000a,Zurita:2002a} and GRS~1915+105
\citep{Neil:2007a}.

Another difference between CVs and LMXBs is that the mechanism for
producing superhump light in CVs, enhanced heating of the disk by
tidal stresses induced by the companion star, should not be important
in LMXBs \citep{Haswell:2001a}. This is because the optical light
should not be dominated by viscous dissipation, but instead by
irradiation by X-rays. This suggests that instead the superhump arises
from coupling of irradiation to tidal distortion of the disk. In
support of this, \citet{Haswell:2001a} show that disk simulations do
produce a $\sim10$\,\%\ increase in disk area at the time of superhump
maximum. \citet{Smith:2007a} note that the period excess in LMXBs is
also smaller than that in CVs and suggest that the precession period
is longer because LMXB disks should be hotter and thicker due to
irradiation.

The case of XTE~J1118+480 is especially interesting. Superhumps were
seen both during outburst, with varying morphology and period
\citep{Uemura:2000a}, and late in the decay to quiescence
\citep{Zurita:2002a}. The superhump periods inferred were just
0.1--0.6\,\%\ of the orbital period. Coupled with the extremely small
mass ratio inferred spectroscopically ($0.037\pm0.007$
\citealt{Orosz:2001a}) this makes XTE~J1118+480 an important object
for studying the extreme limits of superhump behavior in which the 2:1
resonance may become the dominant mechanism for exciting disk
eccentricity.

\subsection{Super-orbital Periods}
\label{WarpSection}

Other periods longer than the orbital period can arise. The most
famous is the 35\,day period seen in the neutron star LMXB Her~X-1.
This manifests in both a long modulation in optical light curve
morphologies \citep{Gerend:1976a} and changes in X-ray pulse profiles.
The optical behavior was modeled successfully with a precessing tilted
accretion disk which modulates shadowing of the companion star on a
35\,day period. The mechanism for inducing this tilt is now believed
to be radiation-driven warping of the accretion disk
\citep{Wijers:1999a}.

\citet{Ogilvie:2001a} investigated the stability of accretion disks to
radiation driven warping in systems of a wide range of mass ratios and
binary separations. They focused on the two lowest order warping
modes, 0 and 1, and identified several regimes where disks were either
stable or unstable to mode-0 perturbations, to mode-1 perturbations,
or to both. Reassuringly, Her~X-1 lies in the region unstable to
mode-0 only, and most other systems in the unstable regime also
exhibit super-orbital periodicities of some form.
\citet{Clarkson:2003a} tested this description more thoroughly with a
dynamic power-spectral analysis of RXTE lightcurves. Systems such as
Her~X-1 and LMC~X-4 which are in the pure mode-0 regime exhibit
relatively stable periodicities. SMC~X-1 which lies close to the
region of mode-1 instability exhibits a less stable long period,
whereas Cyg~X-2 which is unstable to both mode-0 and mode-1
oscillations exhibits complex and multi-periodic behavior.

\subsection{Quasi-Periodic Oscillations}

Quasi-periodic oscillations, or QPOs, are common on many timescales in
LMXBs. Some of the super-orbital periods discussed above would qualify
as long-period QPOs rather than strict periodicities. At the other
extreme of timescale there is a rich phenomenology of high frequency
(milliseconds to seconds) QPO behavior seen in X-ray observations of
LMXBs arising from the inner accretion flow. We will not discuss that
here and focus on those QPOs visible in the optical and/or UV; for a
review of X-ray QPOs see \citet{vanderKlis:2006a}.

A QPO is a repeating signal that is not strictly periodic. It may
wander in frequency or exhibit changes in phase. A transient signal,
for example a decaying series of pulses, will also manifest as a QPO.
In a Fourier transform, a QPO appears as a peak with finite width,
whereas a true coherent periodicity has width limited only by the
time-period sampled by the data. QPOs are often characterized by their
coherence, $Q=\nu/\Delta \nu$, which is a measure of how broad the QPO
is in frequency space. A low coherence QPO is essentially an excess of
noise around a peak frequency, defining a preferred but not unique
timescale.

One class of optical QPOs has been seen in two UCXBs now.
\citet{Chakrabarty:2001a} identified a strong optical and UV QPO
around 1\,mHz in the 42\,min orbital period system 4U~1626--67. This
was seen to be moderately coherent ($Q\sim8)$ stronger in the UV than
optical, and completely absent in X-rays. The authors attributed the
effect to a precessing warp in the inner accretion disk. Subsequently
a similar feature also around 1\,mHz has been seen in 4U~0614+091
(Zhang et al., in preparation). This system is believed to have an
orbital period of 51\,min \citep{Shahbaz:2008b}, so the similarity of
timescales to 4U~1626--67 is striking.

A timescale around 1\,mHz also appears to be significant in quiescent
SXTs, although this is probably coincidence given the very different
orbital periods and physical conditions. \citet{Hynes:2003b}
identified a break in the power-density spectrum of the 8\,hr period
system A\,0620--00 at around 1\,mHz. An actual QPO at 0.78\,mHz was
seen in the 6.5\,day period V404~Cyg \citep{Shahbaz:2003a} and at
2\,mHz in the 4.1\,hr system XTE~J1118+480 \citep{Shahbaz:2005a}. The
favored explanation for QPOs in quiescent SXTs has been that they are
associated with the transition from a thin accretion disk to an
evaporated advective flow, although the interpretation is probably not
as simple as associating the QPO frequency with the Keplerian rotation
period at the transition radius.

\section{Spectroscopy}
\label{SpectroscopySection}

\subsection{Emission and Absorption Line Spectra}
\label{LineSection}

\begin{figure}[t]
\includegraphics[width=2.0in]{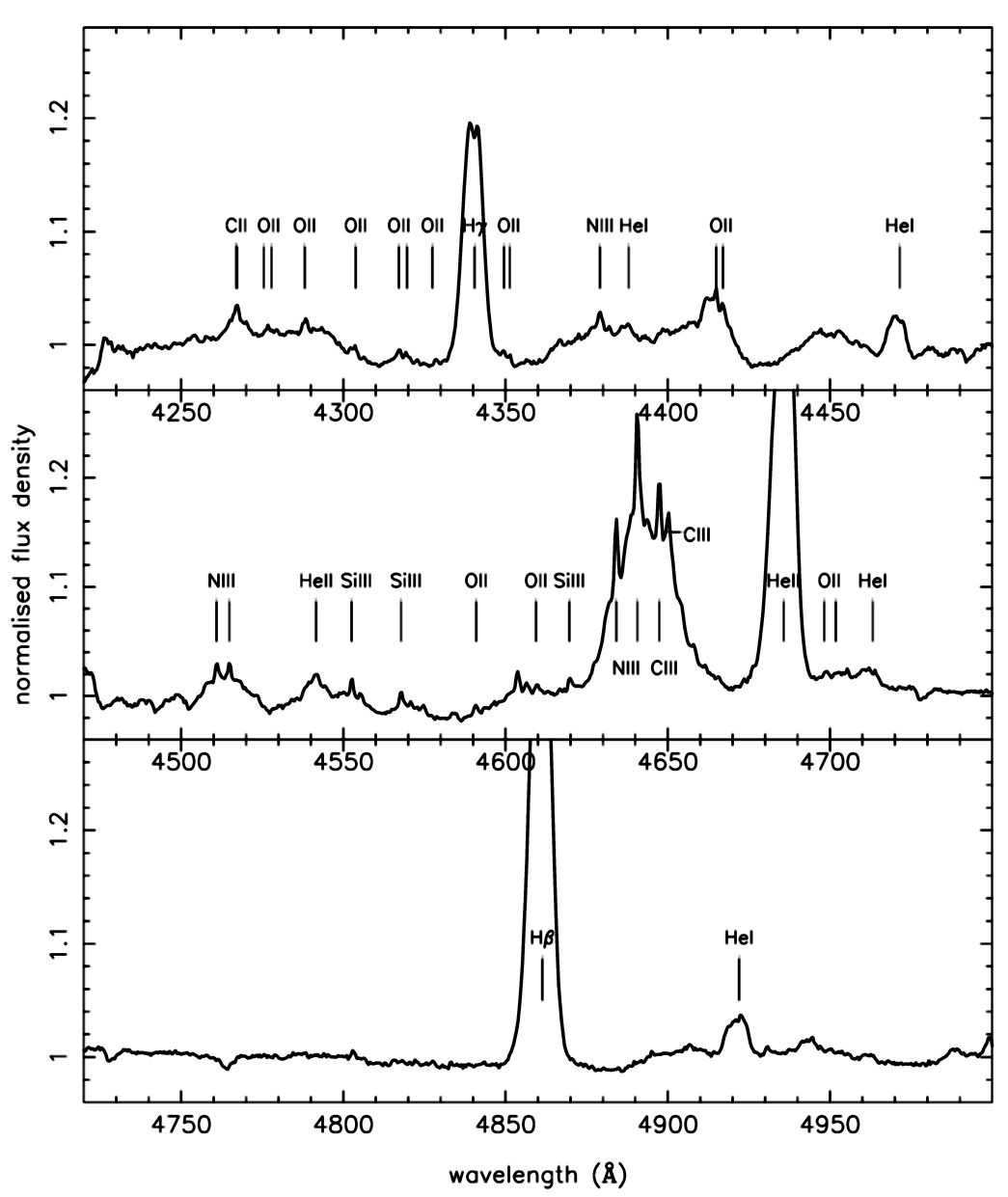}
\includegraphics[width=1.55in]{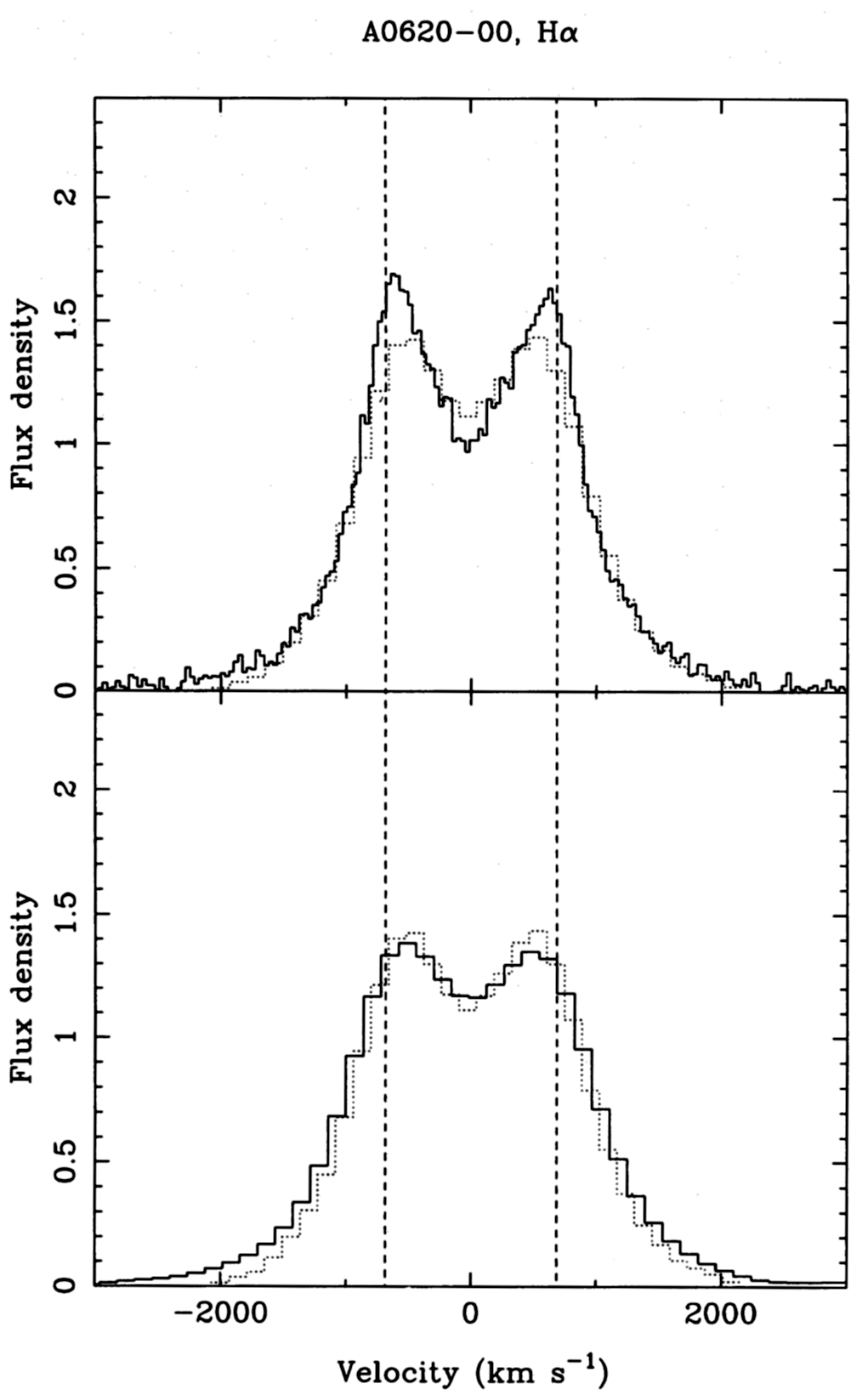}
\caption{Left: Blue spectrum of Sco~X-1 from
  \citet{Steeghs:2002a}.  This shows the typical range of atomic
  emission lines present in the spectrum of a bright LMXB.  Right:
  H$\alpha$ profiles of A\,0620--00 from \citet{Marsh:1994a}.  This
  illustrates the typical double-peaked disk profile especially seen in quiescent
  LMXBs.  Solid
  data are that of Marsh, dotted is from \citet{Johnston:1989a}.  In
  the lower panel, Marsh's data has been binned at blurred to match
  that of Johnston.  Dashed lines indicate the velocity of the outer
  edge of the disk inferred by Marsh. }
\label{SpecFig}
\end{figure}

Luminous X-ray binaries can show a range of emission lines, although
their spectra are nowhere near as rich as those of planetary nebulae
or active galactic nuclei can be. We show an example, the blue
spectrum of Sco~X-1, in Fig.~\ref{SpecFig}.  H\,{\sc i} lines of the Balmer
series in the optical and Brackett series in the infrared are common,
but by no means ubiquitous. They can be absent, and sometimes show
absorption structure. He\,{\sc i} lines often accompany H\,{\sc i}.
More reliably present are He\,{\sc ii} lines in the optical (dominated
by 4686\,\AA) and ultraviolet (1640\,\AA).

Besides hydrogen and helium, the main other lines seen are of carbon,
nitrogen, oxygen, and silicon. In the optical the strongest of these
features is a blend of N\,{\sc iii} and C\,{\sc iii} lines around
4640\,\AA\ commonly referred to as the Bowen blend
\citep{McClintock:1975a}, but many weaker lines are present and can be
seen in high quality spectra of Sco~X-1 \citep{Steeghs:2002a}. In the
UV, several strong resonance lines dominate: C\,{\sc iv}
1548,1551\,\AA, N\,{\sc v} 1239,1243\,\AA, Si\,{\sc iv}
1394,1403\,\AA, and O\,{\sc v} 1371\,\AA\ are the most prominent.
Dramatic differences in the relative intensities of these have been
seen between objects with carbon and oxygen lines sometimes completely
absent. This has been interpreted as evidence for CNO processing in
the donor star \citep{Haswell:2002a}. Among ultracompact systems,
hydrogen, and sometimes even helium may be absent leaving spectra
dominated by carbon and oxygen \citep{Nelemans:2004a,Nelemans:2006a}.

In quiescence, higher excitation lines of He\,{\sc ii} together with
the CNO lines are absent, and only H\,{\sc i} and very weak He\,{\sc
  i} are present in emission. Provided the disk is sufficiently dim,
the photospheric absorption spectrum of the companion emerges
facilitating measurement of its radial velocity curve and rotational
broadening and derivation of the system parameters. This will be
discussed in Section~\ref{ParameterSection}.

\subsection{Emission Line Profiles and Doppler Tomography}

\begin{figure}[t]
\includegraphics[width=3.5in]{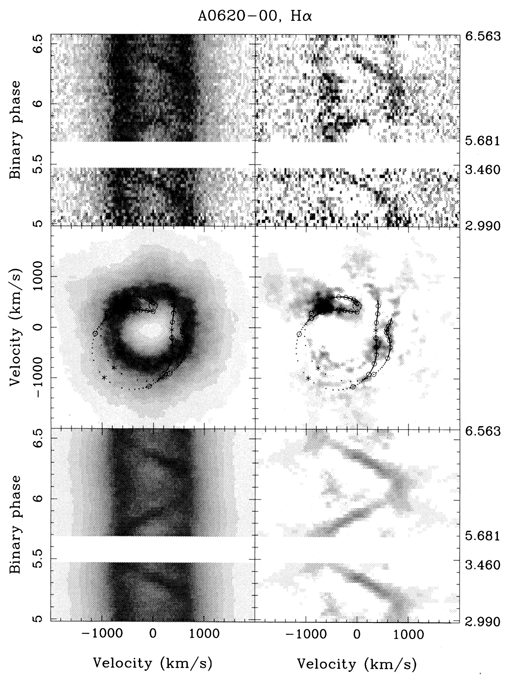}
\caption{Doppler tomogram of H$\alpha$ emission in A~0620--00 from
  \citet{Marsh:1994a}. The right hand panels shows the effect of
  subtracting the non-varying part of the trailed spectrum (equivalent
to the symmetric part of the tomogram. The upper panel shows the data,
the middle is the tomogram, and the lower is a reconstruction of the
data from the tomogram.}
\label{MarshFig}
\end{figure}

Quiescent LMXBs present the simplest emission line profiles to
understand. Optical hydrogen and helium lines are seen with a
double-peaked profile (e.g.\ Fig.~\ref{SpecFig}) consistent with
expectations from an approximately axisymmetric Keplerian accretion
disk as also seen in CVs \citep{Horne:1986a}. The blue wing of the
profile comes from the side of the disk approaching us, and the red
wing from that which is receding. Most sensitively, the peak
separation can measure the velocity at the outer edge of the disk.
More close examination, however, reveals changes in the line profiles,
with the relative strengths of the two peaks varying. To examine how
this relates to the binary orbital phase, it is usual to plot a
trailed spectrogram, with wavelength (equivalent to velocity for a
single line) on the x-axis and orbital phase on the y-axis. Good
quality data then reveals that the asymmetries arise from a third peak
moving back and forth sinusoidally, tracing out an S-wave in the
trailed spectrum (e.g Fig.~\ref{MarshFig}; \citealt{Marsh:1994a}).

To better understand how such components can be associated with
locations in the binary, the technique of Doppler tomography has been
developed. A number of documents have reviewed this well, for example
\citet{Marsh:2001a}. The key idea is that any given component of the
binary should have a defined velocity in the orbital plane that can be
resolved into $x$ and $y$ components, $V_x$ and $V_y$. Each such
component will also lead to a single S-wave in a trailed spectrogram.
Doppler tomography is essentially a transformation from visualizing
the data in terms of radial velocity and phase to an alternative
visualization in terms of $V_x$ and $V_y$. Thinking of the $V_x - V_y$
plane in terms of polar coordinates, the phasing of an S-wave
component dictates the azimuth of a spot in the $V_x - V_y$ plane, and
its amplitude determines the distance of the point from the origin
($V_x = V_y = 0$). Doppler tomography can also be thought of as
representing the trailed spectrogram as a sum of an arbitrary number
of S-waves.

The dominant component that is expected in a Doppler tomogram is a
ring of emission in velocity space corresponding to the accretion disk
(see e.g.\ Fig.~\ref{MarshFig}; \citet{Marsh:1994a}). Since this is
represented in velocity space the inner disk has high velocities and
so is on the outside of the ring in the tomogram; in this sense
tomograms are inside-out compared to visualizing things in real space.
The other component most often seen is a structure on the left hand
side corresponding to the accretion stream, the stream-impact point,
or a bulge in the disk downstream of it. In quiescent LMXBs such as
A~0620--00 this is relatively well behaved and lies at velocities
expected for the accretion stream \citep{Marsh:1994a}. In persistent
LMXBs, however, the maximum emission often lies below the expected
stream velocities in the tomogram \citep{Casares:2003a,Pearson:2006a}
and may even appear in the lower-left quadrant \citep{Hynes:2001a}.
This is usually attributed to a bulge introduced by the stream-impact
or material overflowing the disk, but its origin is not yet fully
understood. The final component sometimes revealed in some emission
lines is the companion star. This will be discussed more in
Section~\ref{BowenSection}.

There are several limitations of the technique to keep in mind,
especially as X-ray binaries can violate a number of these. Doppler
tomography assumes that emission lines are optically thin and that all
of the flux is visible all of the time. This requires lines that do
not modulate in integrated intensity over the binary orbit, yet
observationally such modulations usually are present, with eclipses
presenting the most dramatic example. The new technique of modulation
tomography \citep{Steeghs:2003a} partially addresses this by allowing
the line intensity to modulate sinusoidally. This technique has been
successfully applied to the quiescent LMXB XTE~J1118+480 by
\citet{Calvelo:2009a}, where it is found that the disk component, as
expected, does not modulate in intensity but the accretion stream
impact point does. Another limitation which can more justifiably be
assumed satisfied is that all motions are in the plane (i.e.\ there is
no $V_z$ component). Even so, a disk wind, for example, violates this
assumption and wind emission would confuse the reconstruction. A final
thing to keep in mind, but not a limitation of the technique as such,
is that a Doppler tomogram is reconstructed in velocity space, not
real space. The velocities of emitting material can be constrained,
but additional assumptions, such as system parameters, are required to
associate these velocity components with physical components of the
binary. Conversely, of course, this means that Doppler tomograms can
be used to constrain those system parameters, and we will return to
this idea in Section~\ref{BowenSection}. Even knowing system
parameters, the mapping between real and velocity space is not
one-one, and so there may be ambiguity about the physical location of
emitting material.

\subsection{Mass Determinations in Quiescent Systems}
\label{ParameterSection}

Binary system parameters, and especially masses, are essential for
many purposes. Specific parameters provide information about the
evolutionary state of a system, while distributions in period and
compact object mass, for example, test models for the formation and
evolution of binary populations \citep{Fryer:2001a,Pfahl:2003a}. A
compact object mass in excess of 3\,M$_{\odot}$ is considered the most
convincing evidence for a black hole, while precise measurements of
neutron star masses could constrain the equation of state of neutron
star matter (e.g. \citealt{Ozel:2009a}). Finally, most of the
observational techniques described here depend, to a greater or lesser
degree, on knowing the system parameters for a quantitative
interpretation.

Dynamical parameter determinations are underpinned by Kepler's Third
Law which can be expressed as:
\begin{equation}
f(M)=\frac{K_2^3P_{\rm orb}}{2\pi G}=M_1\frac{\sin^3 i}{(1+q)^2}
\end{equation}
where $K_2$ is the radial velocity semi-amplitude of the donor star,
$P_{\rm orb}$ is the binary orbital period, $M_1$ is the compact
object mass, $i$ is the binary inclination, and $q=M_2/M_1$ is the
mass ratio. $f(M)$ is termed the mass function. It is of crucial
importance because a) it can be measured from the radial velocity
curve alone ($K_2$ and $P_{\rm orb}$) and b) it provides a strict
lower limit on $M_1$ (since $\sin i \leq 1$ and $1+q > 1$). Many
objects have been classified as a black hole purely on the basis of
mass functions greater than the assumed maximum mass of a neutron star
around 3\,M$_{\odot}$. The most convincing example, and an excellent
case-study for system parameter determination in general, is V404~Cyg
\citep{Casares:1992a,Casares:1994a}. It has an orbital period of
6.47\,days and $K_2=208.5\pm0.7$\,km\,s$^{-1}$. This leads to a mass
function $f(M)=6.08\pm0.06$\,M$_{\odot}$.

Beyond this, it can be seen that determination of the actual compact
object mass requires measurement of the binary mass ratio and
inclination. The preferred method to estimate the mass ratio is to
observe rotational broadening in photospheric absorption lines from
the donor star. Since the projected radial velocity semi-amplitude,
$K_2$, and the rotational broadening, $v \sin i$, are both affected by
orbital inclination in the same way, the ratio of $v \sin i / K_2$ is
purely a function of the ratio of the radius of the donor star to the
binary separation and hence is a function of $q$ only (although
details such as limb darkening are important for precise measurement
of $v \sin i$ from observed rotationally broadened lines). A good
approximation (see e.g.\ \citealt{Wade:1988a}) is
\begin{equation}
V_{\rm rot} \sin i = 0.462 K_2 q^{1/3} (1+q)^{2/3}.
\end{equation}

In practice, rotational broadening measurements are challenging since
the expected values, typically below $100$\,km\,s$^{-1}$, are often
close to the spectral resolution of the data used. The usual method to
extract this information is to observe both the target and a template
star of matched spectral type and then convolve the template spectrum
with rotational broadening profiles until an optimal match is found.
In the case of V404~Cyg, \citet{Casares:1994a} measured $V_{\rm rot}
\sin i = 39.1\pm1.2$\,km\,s$^{-1}$ which, together with $K_2$ cited
above, yields $q=0.060^{+0.004}_{-0.005}$.

An alternative approach to the mass ratio is to attempt to measure the
radial velocity semi-amplitude of the compact object, $K_1$, from disk
emission lines, since $q=K_1/K_2$. In practice this has been fraught
with errors, with emission line radial velocity curves exhibiting
incorrect phasings and different systemic velocities to the companion
star. It is argued that these effects arise from asymmetries far out
in the disk, and that they can be minimized by measuring just the high
velocity wings of a line, but even then, mass ratios from rotational
broadening are preferable.

The preferred method to determine the inclination would be to model
eclipses (see Section~\ref{EclipseSection}). Most systems (including
all Galactic black hole candidates) do not eclipse, however, so
instead the usual method has been to measure ellipsoidal variations
(see Section~\ref{EllipsoidalSection}). Continuing our case-study of
V404~Cyg, the ellipsoidal variation study of \citet{Shahbaz:1994a} is
illustrated in Fig.~\ref{ShahbazFig}. These authors model IR
ellipsoidal modulations to deduce $i=56\pm4^{\circ}$. Combined with
the mass function and mass ratio measurements discussed above, this
implies $10\,{\rm M}_{\odot} < M_1 < 15\,{\rm M}_{\odot}$ with a
preferred value of $M_1=12$\,M$_{\odot}$. As discussed in
Section~\ref{EllipsoidalSection}, an important concern with
ellipsoidal studies is that the modulations may be diluted by disk
contamination. \citet{Shahbaz:1996a} measured this in the IR and
placed an upper limit on the disk contribution of 14\,\%. This at most
reduces the derived mass by 2\,M$_{\odot}$ from 12\,M$_{\odot}$ to
10\,M$_{\odot}$.

It should be emphasised that we have focused on V404~Cyg because it is
both one of the best studied systems, and one with the cleanest
results. In most quiescent SXTs, measurements are less precise and
concerns about disk contamination are more serious, and consequently
most mass determinations in these systems are much less secure. This
method has now been applied to about 15 quiescent SXTs. For a recent
compilation, see \citet{Casares:2007a}.

\subsection{Mass Determinations in Luminous Systems}
\label{BowenSection}

As the accretion rate increases from quiescence, the optical light
quickly becomes dominated by the heated accretion disk and companion
star. This means that in most persistently luminous LMXBs, we never
have the opportunity to perform a radial velocity study of
photospheric absorption lines from the unheated portions of the
companion. There are a few exceptional systems in which the donor star
is a giant and is still visible even in luminous states. Cyg~X-2 is
the most famous example \citep{Casares:1998a} and 2S~0921--636 is
another \citep{Shahbaz:2004a,Jonker:2005a}.

In other systems we must seek a different source of radial velocity
information. A new approach was suggested by observations of Sco~X-1
which revealed narrow emission components in the Bowen blend (see
Section~\ref{LineSection}) that moved in anti-phase with broader
components attributed to the disk \citep{Steeghs:2002a}. The radial
velocity curve of just the sharp components yielded
$K_2>77$\,km\,s$^{-1}$ and the authors derived a neutron star mass
$\sim1.4$\,M$_{\odot}$, consistent with that usually expected. The
major caveat here is that the emission lines would originate from the
heated inner face of the donor star, and so their center-of-light
would not coincide with the donor's center of mass. This is why the
$K_2$ value derived by this method is only a lower-limit on the true
value. To improve on this requires modeling of the binary geometry to
attempt to estimate by how much $K_2$ is underestimated
\citep{MunozDarias:2005a}.

Sco~X-1 was an ideal case. Only one other system, the long-standing
black hole candidate GX\,339--4, has shown clear, sharp, moving
N\,{\sc iii} and C\,{\sc iii} components in individual spectra
\citep{Hynes:2003c}. These observations provided both a convincing
orbital period of 1.76\,days, and a mass function of
$f(M)=5.8\pm0.5$\,M$_{\odot}$, finally confirming the black hole
nature of this source. In other objects, such as 2A~1822--371, sharp
components cannot be identified in the line profiles directly, but the
companion star can be picked out very effectively using Doppler
tomography \citep{Casares:2003a}. For a more complete review of what
has been achieved by this method, see \citet{Cornelisse:2008a}.

\section{Rapid Variability}
\label{RapidSection}

\subsection{Echo-Mapping}

Multiwavelength variability is a near universal characteristic of
X-ray binaries. X-rays vary due to rapid changes in the inner
accretion flow on timescales of milliseconds and longer. These X-rays
then irradiate the outer accretion disk and companion star, resulting
in reprocessed optical and UV radiation which is expected to be
imprinted with the same variability as the X-ray signal. An important
difference, however, is that the optical emission and X-rays originate
from a volume of significant spatial extent, resulting in light travel
time delays between the X-rays and the reprocessed emission. It is
then possible to infer information about the geometry and scale of the
reprocessing region from the lags measured between X-ray and
optical/UV variability; this technique is known as reverberation or
echo-mapping, as the reprocessed light behaves as an echo. Echo or
reverberation mapping is not uniquely applied to X-ray binaries. Much
of the development and application of the technique has been for
active galactic nuclei (AGN); see for example \citet{Peterson:2006a}
for a recent review of the AGN problem, and \citet{OBrien:2002a} for
the application to X-ray binaries.

The key idea is that local optical (or UV) variability, in either
lines or continuum, is induced by reprocessing of X-ray variability,
but with a lag varying with reprocessing location. Each such location
can be thought of as responding to X-rays with a delta-function
response at a delay time determined by the path difference between
direct and reprocessed emission. The total optical response is then
the sum of lagged responses from all the reprocessing elements. For a
delta-function variation in the X-rays the optical response is then
termed the transfer function, and measures how strong the response is
as a function of the delay, effectively encoding information about the
reprocessing geometry. For continuously variability, the optical
lightcurve can be modeled as a convolution of the X-ray lightcurve
with the transfer function:
\begin{equation}
L_{\rm opt}(t) = L_{\rm X} * \Psi = \int L_{\rm X}(t-\tau)\Psi(\tau)d\tau
\label{TransferFunctionEquation}
\end{equation}
where $L_{\rm X}$ and $L_{\rm opt}$ are the X-ray and optical
luminosities, $\Psi$ is the transfer function, and $\tau$ is the X-ray
to optical lag.

Several assumptions are inherent in this description. It is assumed
that the optical responds linearly to the X-rays, or at least that a
non-linear response can be linearized for small perturbations. It is
also implicit that the X-rays originate from a point source, or at
least a region much smaller in spatial extent than the reprocessing
region. Finally, to determine geometric information it is necessary
that the lags be geometric in origin; significant reprocessing times
would compromise this.

\subsubsection{Geometrical modeling of the response}

\begin{figure}[t]
\includegraphics[angle=180,width=3in]{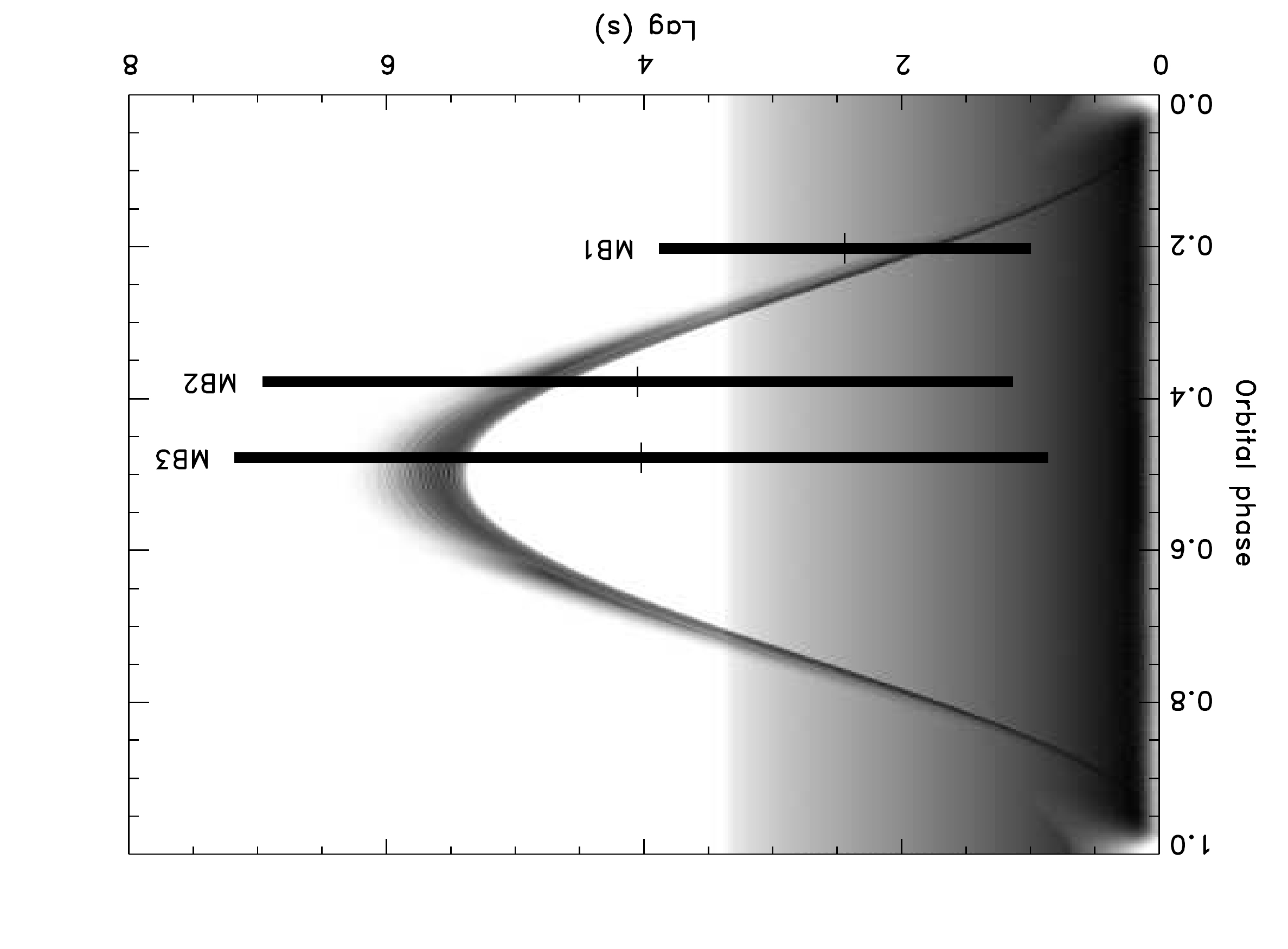}
\caption{Strength of response (indicated by grayscale) as a function
of orbital phase and lag, as calculated by the binary modeling code
of \citet{OBrien:2002a}.  Solid bands indicate the ranges of lags
inferred from X-ray bursts by \citet{Hynes:2006a} for EXO~0748--676.}
\label{EchoTomFig}
\end{figure}

In the case of X-ray binaries, we have a clearer expectation of the
reprocessing geometry than in AGN. We anticipate reprocessing from the
accretion disk around the compact object, possibly enhanced at a bulge
where material feeds into the disk from the companion star. We also
might expect some reprocessing from the heated inner face of the
companion star. \citet{OBrien:2002a} and \citet{MunozDarias:2005a}
modeled the reprocessing geometry to predict transfer functions for a
variety of binary parameters. An example as a function of orbital
phase is shown in Fig.~\ref{EchoTomFig}. Simplistically, one expects
two components. The disk will extend from zero lag to $r_{\rm disk}(1
+ \sin i)$ where $r_{\rm disk}$ is the disk radius in light seconds
and $i$ the binary inclination. Within this range the shape of the
response is strongly sensitive to the inclination and somewhat less so
to the degree of disk flaring. The response from the companion star
approximately oscillates within the range $a(1 \pm \sin i)$ over the
course of the binary orbit, where $a$ is the binary separation in
light seconds. The strength and width of the companion response is a
strong function of the mass ratio and disk thickness (which determines
how much of the companion is shielded). One of the great appeals of
applying echo-mapping to X-ray binaries is that with phase-resolved
observations of the companion echo over the orbit, one could measure
both $a$ and $i$ independently of other techniques and assumptions.

\subsubsection{Empirical descriptions of the response}

Many echo-mapping studies use more pragmatic, and less model-dependent
approaches. Ideally, one would take high quality X-ray and optical
lightcurves and deconvolve them to directly determine the shape of the
transfer function. This is the basis of the maximum entropy
echo-mapping technique \citep{Horne:1994a} in which a maximum entropy
regularization method is used to suppress the problem of fitting the
noise. Unfortunately, typical X-ray binary datasets do not have the
signal-to-noise ratio for this to work effectively.

A simpler approach has been developed in which a very simple
functional form is adopted for the transfer function, either a
rectangular response \citep{Pedersen:1982a} or a Gaussian
\citep{Hynes:1998a}. Both are introduced as approximations to the
response rather than for a physically motivated reason, and amount to
the assumption that the data only constrain the mean lag and the
amount of smearing out in time of the variability. The Gaussian
formulation essentially yields the first two moments of the delay
distribution. There is a potential difficulty with this approach as
noted by \citet{MunozDarias:2007a}. If the X-ray lightcurve is
convolved with a Gaussian this smooths the data and has the effect of
suppressing the noise. If the noise in the X-ray lightcurve is
significant compared to the real variability (or if there is
additional variability that does not correlate with that in the
optical) then the $\chi^2$ of a fit may be reduced by adopting an
artificially high Gaussian width, i.e.\ over-smoothing the data. Thus
there is a possibility that the widths of transfer functions derived
in this way may be over-estimated.

\begin{figure}[t]
\includegraphics[width=2.2in]{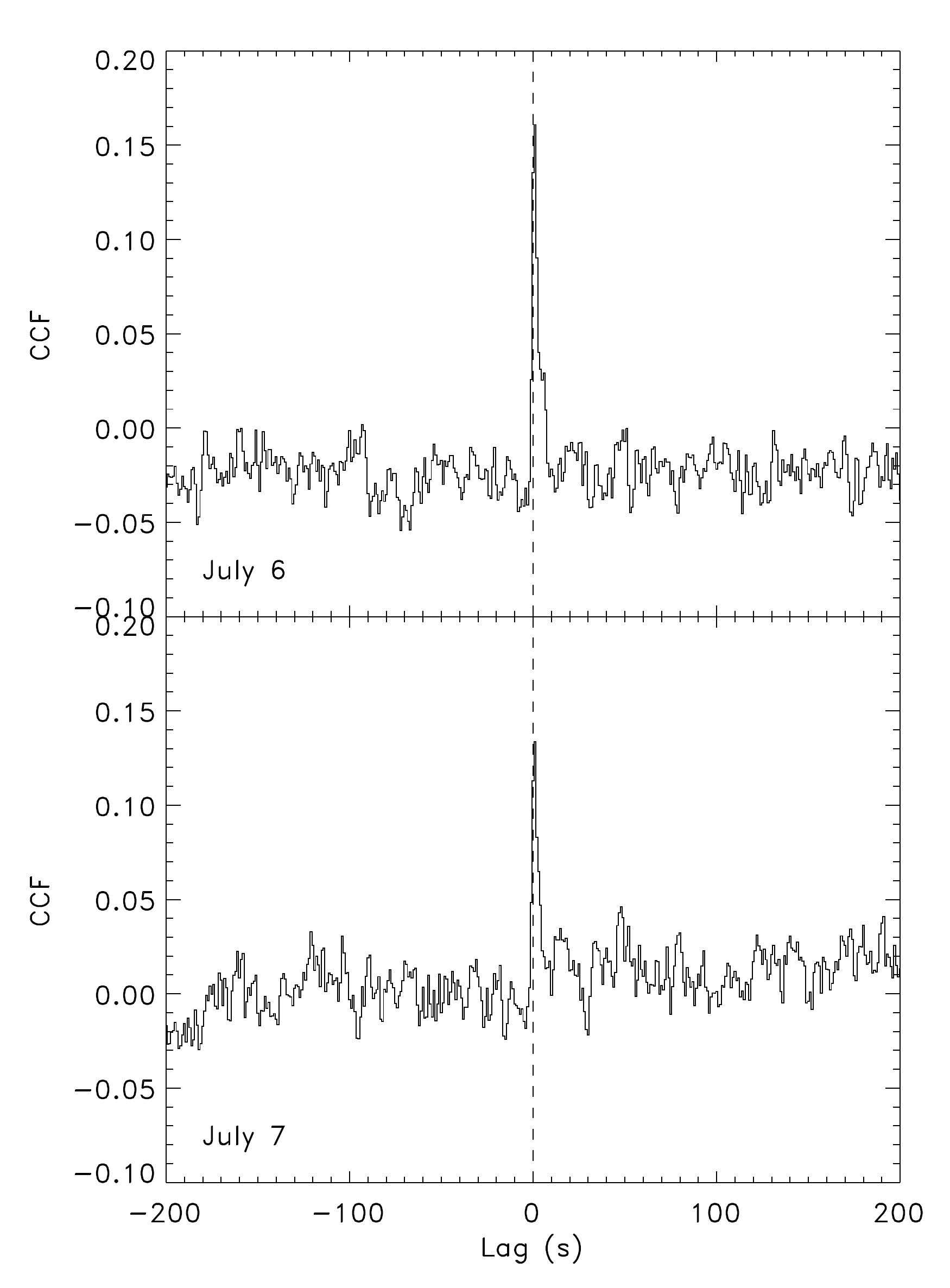}
\includegraphics[width=2.0in]{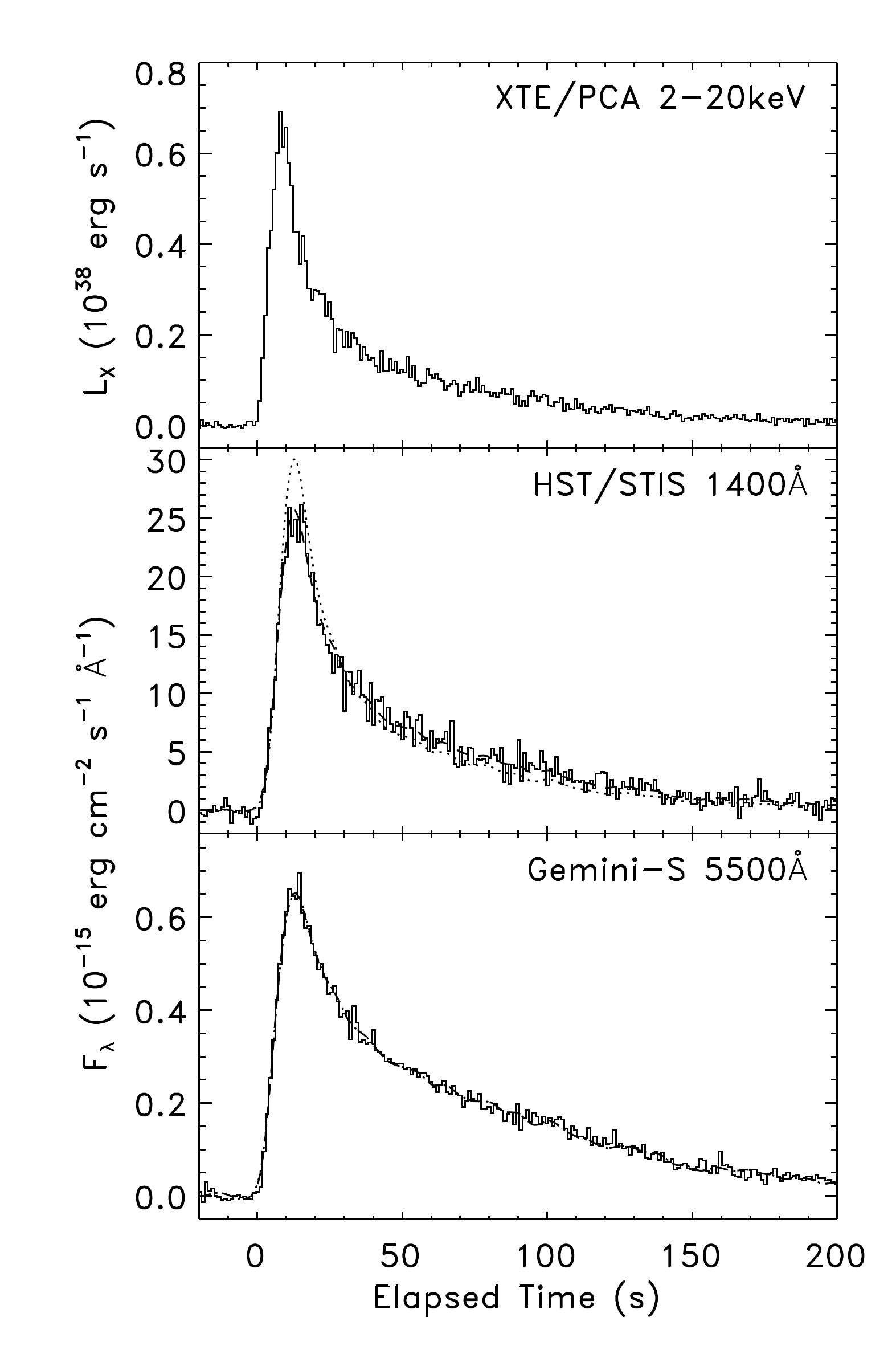}

\caption{Left: X-ray/optical cross-correlations of Swift~J1753.5--0127
  early in outburst, from \citet{Hynes:2009b}. Right: Simultaneous
  burst profiles in EXO~0748--676 from \citet{Hynes:2006a}. Dotted
  lines assume fixed normalization of the UV flux relative to the
  optical, solid line fits allow this normalization to float.}

\label{CCFFig}
\label{BurstFig}
\end{figure}

An even simpler, but widely used approach is to measure the
cross-correlation function (CCF) of the two datasets, defined for
continuously sampled data as:
\begin{equation}
{\rm CCF} (\tau) = \frac{\int[f(t)-\bar{f}][g(t+\tau)-\bar{g}]}{\sigma_f \sigma_g}
\end{equation}
where $f$ and $g$ are the driver and echo time-series respectively,
$\bar{f}$ and $\bar{g}$ are their means, $\sigma_f$ and $\sigma_g$ are
their standard deviations, and $\tau$ is the lag at which the CCF is
being evaluated. \citet{Gaskell:1987a} and \citet{Edelson:1988a}
discuss two approaches to implementing this calculation for discretely
and unevenly sampled data. This yields an estimate of the mean lag but
information about the smearing is hard to extract, and care should be
taken in interpreting the results of a cross-correlation analysis
\citep{Koen:2003a}. If the driver-echo relationship is accurately
described by the convolution in
equation~\ref{TransferFunctionEquation} then the CCF is equivalent to
the driver auto-correlation function (ACF) convolved with the transfer
function. In cases where the driver ACF is relatively narrow, the CCF
may provide a reasonable approximation to the transfer function. We
show an example in Fig.~\ref{CCFFig}, based on X-ray/optical data from
the black hole transient Swift~J1753.5--0127 \citep{Hynes:2009b}.

\subsection{Type I X-ray Bursts}

Type-I X-ray bursts are thermonuclear explosions on the surface of a
neutron star in an LMXB \citep{Strohmayer:2006a}. They represent an
enormous increase in the X-ray flux, a factor of twenty or more in
lower luminosity systems, rising on a timescale of a few seconds.
X-ray bursts then have the potential to be an ideal echo-mapping
probe. Hence it is using these events that some of the first
echo-mapping experiments were performed. Reprocessed optical bursts
were discovered in the late 1970's in the LMXBs 4U~1735--444 and Ser
X-1 \citep{Grindlay:1978a,McClintock:1979a,Hackwell:1979a}. The
optical flux was found to rise by nearly a factor of two and lag a few
seconds behind the X-rays. It was immediately appreciated that the
optical flux was several orders of magnitude too high to be due to
direct emission from the neutron star surface, and hence that the
brightening must be due to reprocessing of X-rays by the much larger
projected area of the accretion disk and/or companion star. The 2.8\,s
lag in 4U~1735--444 \citep{McClintock:1979a} supported this
interpretation, being consistent with the expected light travel time
delays in this short-period binary.

The large amplitudes of X-ray bursts drive a stronger optical response
than X-ray flickering does, and so provide additional information not
available when variability is a small perturbation. Observations never
record the total bolometric luminosity, but always that within a
specific bandpass. Consequently the shape of the observed reprocessed
lightcurve depends on the spectral evolution as the reprocessor cools
and the peak of the spectrum moves into or out of the bandpass used.
Shorter wavelengths are sensitive to hotter material, they are
expected to decay more rapidly, and hence multicolor observations of
X-ray bursts provide some temperature sensitivity, in addition to that
lag information which is available even for smaller perturbations.
This is not a subtle effect, and the early observations indicate the
reprocessor temperature typically doubles during a burst
\citep{Lawrence:1983a}.

The best dataset for the method yet obtained was obtained for
EXO~0748--676 \citep{Hynes:2006a}. Four bursts were observed over two
successive nights using RXTE and Gemini-S, providing some phase
information, and one of these was also observed at high
time-resolution in the far-UV by {\it HST}/STIS (Fig.~\ref{BurstFig}).
The latter was a unique observation to date, providing far more
sensitivity to high temperature responses than is possible with
optical data alone, and also yielding a time-resolved UV spectrum,
facilitating a direct test of the expectation that the reprocessed
light should be close to a black body. Perhaps most interestingly, the
three bursts observed sampled enough of a phase range to see apparent
changes in the lag and smearing as a function of orbital phase
(Fig.~\ref{EchoTomFig}). The phasing and amplitude of the changes seen
are both consistent with expectations of models in which both the disk
and the companion star contribute to reprocessing. This is one of the
few observations to date that can claim to be true echo-tomography
exploiting different viewing angles of the binary.


\subsection{Flickering in Persistent Neutron Star Systems}

While bursts provide an ideal signal for echo-mapping in some neutron
star binaries, we must seek another technique in black hole systems
and non-bursting neutron stars. An alternative source of variability
is provided by the flickering that seems a ubiquitous signature of
accretion. Since this flickering is always present at some level,
unlike bursts which only recur every few hours, flickering variability
can potentially provide phase-resolved information in any system. This
potential has yet to be fully realized, however. It has been found
that the optical response is rather weak, with standard deviations of
only a few percent in the optical lightcurves. Consequently, high
signal-to-noise observations are needed to pick out a measurable
correlation. Even then, success is typically only achieved when high
levels of variability are present, with other datasets yielding a
non-detection.

There are several bright and persistently active neutron star X-ray
binaries that potentially provide ideal targets for these studies. It
is in fact somewhat surprising that more has not yet been achieved
with these. The bright neutron star system Sco X-1 was observed by
\citet{Ilovaisky:1980a} and \citet{Petro:1981a}. Both found
correlations, with evidence for lags and substantial smearing of the
response; \citet{Petro:1981a} described the optical response as a
low-pass filtered version of the X-rays, with variability on
timescales $<20$\,s smoothed out. \citet{McGowan:2003a} reanalyzed
these datasets with the Gaussian transfer function method. In some
cases, no good fit could be obtained. The pair of lightcurves where
the method did appear to succeed yielded a lag of $8.0\pm0.8$\,s and
Gaussian dispersion of $8.6\pm 1.3$\,s. For comparison, lags of up to
4--5\,s are expected from the disk and 10\,s from the companion star.

\begin{figure}[t]
\includegraphics[angle=180,width=3.0in]{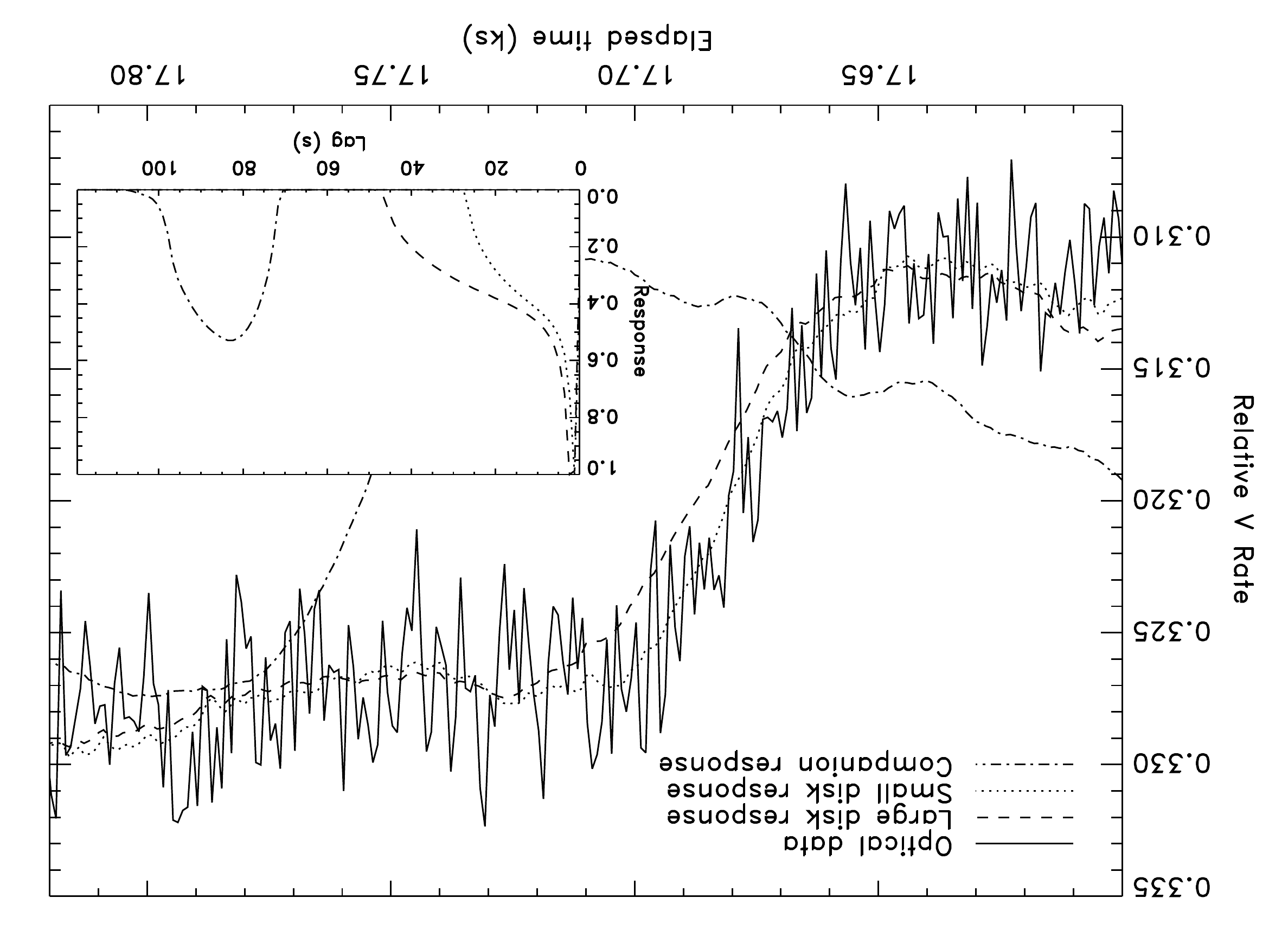}
\caption{Fit to optical data for Cyg~X-2 in the flaring branch.  The
dashed line shows the response from the whole disk, the dotted line is
that from just the inner disk, and the dash-dot line is from 
the companion star.}
\label{CygFig}
\end{figure}

We recently obtained some superb quality observations of Cyg~X-2. Very
clear correlations were seen when the source was on the flaring
branch. A Gaussian transfer function analysis of the whole lightcurve
proved unsatisfactory as we see not only a modulation of the optical
light that is absent in X-rays, but also an apparent variation in the
efficiency of reprocessing from one event to the next. Much more
success was achieved by analyzing individual events in the lightcurve
independently. A Gaussian fit to the transfer function then suggests
lags around 10\,s, as might be expected from the accretion disk in
this long-period (9.8\,day) binary. The data quality are sufficient to
permit direct comparison with model transfer functions as well. We
show in Fig.~\ref{CygFig} a small segment of the optical lightcurve
with fits generated by convolving the X-ray lightcurve with model
transfer functions generated using the code of \citet{OBrien:2002a},
and parameters from \citet{Casares:1998a} and \citet{Orosz:1999a}. The
relative contribution of reprocessing from disk and companion star is
highly sensitive to the amount of shielding of the companion by the
disk, hence we have considered the two components separately. The data
appear much more consistent with the pure disk model than the pure
companion star model suggesting that the disk (and indeed just the
inner disk) dominates the response, at least in the optical continuum.

\citet{MunozDarias:2007a} tried a different approach on Sco~X-1. As
discussed in Section~\ref{BowenSection}, we have realized in recent
years that the Bowen blend of N\,{\sc iii} and C\,{\sc iii} lines
around 4640\,\AA\ often contain a strong component from the companion
star. They used a narrow-band filter encompassing the Bowen blend and
He\,{\sc ii} 4686\,\AA\ to attempt emission line echo-mapping. They
report results from several flaring branch data segments showing
strong correlated variability. As expected, they find longer lags from
the narrow-band filter containing the Bowen blend than from a
continuum bandpass, suggesting that the Bowen response contains a
larger contribution from the companion star. After attempting to
remove the continuum contribution from the Bowen lightcurves, they
estimate a line lag of $\tau_0=(13.5\pm3.0)$\,s and a continuum lag of
$\tau_0=(8.5\pm1.0)$\,s. The difficulty of continuum subtraction with
narrow-band photometry is a limitation of this approach and we can
expect a significant improvement with rapid spectroscopy with new
instrumentation such as ULTRASPEC \citep{Dhillon:2008a}.


\subsection{Another Mechanism for Correlations?}

Fundamental to echo-mapping is the assumption that correlated X-ray
and optical/UV variability indicate reprocessing of the X-rays by
relatively cool material.  We should not take this for granted,
however, and there are some observations which seriously challenge
this assumption.

The first indication of difficulties came from fast optical
observations of the black hole binary GX\,339--4
\citep{Motch:1982a,Motch:1983a}. Dramatic optical variability was seen
extending to extremely short timescales (10--20\,ms), much shorter
than the light travel timescales expected, or the smearing typically
observed in other systems described above. \citet{Fabian:1982a} argued
that the flares most likely originated in cyclotron radiation, with a
brightness temperature $>9\times10^8$\,K. Correlations were seen in a
short (96\,s) simultaneous observation, but of a puzzling nature. The
X-ray and optical were anti-correlated with optical dips apparently
preceding the X-rays by a $2.8\pm1.6$\,s. The connection between X-ray
and optical behavior was further reinforced by the presence of
quasi-periodic oscillations at the same frequency in both energy
bands. The brevity of the simultaneous observation, and the ambiguity
in the lags introduced by quasi-periodic variability left this result
tantalizing however.

New light was shed on this behavior by the 2000 outburst of the black
hole system XTE~J1118+480. A much larger time-resolved database was
accumulated on this object including both simultaneous X-ray/optical
data \citep{Kanbach:2001a} and independent multi-epoch simultaneous
X-ray/UV observations \citep{Hynes:2003a}. Large amplitude X-ray
variability was present, and accompanied by correlated optical and UV
variations. In this case, a positive correlation with the optical/UV
lagging the X-rays was clearly present, leading to hopes that this
would be an ideal echo-mapping dataset. There were serious problems
with this interpretation, however. These were most pronounced in the
optical data \citep{Kanbach:2001a} and included an optical
auto-correlation function narrower than that seen in X-rays, and a
cross-correlation function containing a marked ``precognition dip''
before the main peak. The latter could be interpreted in terms of
optical dips leading X-ray flares by a few seconds, as suggested in
GX\,339--4, suggesting a common origin. Neither of these effects are
expected in a reprocessing model. Light travel times should only act
to smooth out optical responses, and hence broaden the optical
auto-correlation function. Also, the continuum responses (as
considered here) should generally be positively correlated with the
X-rays, not anti-correlated. As in GX\,339--4 the variability extended
to very short timescales ($<100$\,ms) and hence \citet{Kanbach:2001a}
estimated a minimum brightness temperature of $2\times10^6$\,K. They
also suggested that the strange variability properties were the result
of optical cyclosynchrotron emission. These properties become weaker
at shorter wavelengths \citep{Hynes:2003a}, as does the variability,
as might be expected if the behavior originates from a very red source
of emission like synchrotron.

Dominant synchrotron emission in this system was not uniquely
suggested by the variability properties. The very flat UV to near-IR
spectrum had previously been attributed to synchrotron emission
\citep{Hynes:2000a} and the broad-band spectral energy distribution
has been successfully accounted for using a simple jet model
\citep{Markoff:2001a}. Striking support for a synchrotron origin for
the variability has been provided by time-resolved IR observations
during the 2005 outburst of XTE~J1118+480 \citep{Hynes:2006b}.
Simultaneous 2\,s images in the IR $J$, $H$, and $K$ filters could be
used to isolate the color of the IR variability. It was found to be
very red ($F_{\nu} \propto \nu^{-0.78}$), consistent with optically
thin synchrotron emission.

These arguments together provide strong evidence that a jet, or at
least some kind of outflow, is responsible for much of the IR and even
optical emission in XTE~J1118+480 and for the correlated variability.
By extension, the same interpretation may apply to other objects
showing similar properties. GX\,339--4 is of course a prime candidate
and recent more extensive observations of this object show similar
behavior \citep{Gandhi:2008a}. A third object has been added to the
sample in Swift~J1753.5--0127 \citep{Durant:2008a,Hynes:2009b}. Here
early observations near the outburst peak supported a simple
reprocessing interpretation (Figure~\ref{CCFFig}), with a transfer
function consistent with the inferred system parameters, but later
observations showed both negative and positive correlations rather
similar to those in XTE~J1118+480. The full explanation for the
observed correlations and anti-correlations remains to be established,
but this case illustrates that new astrophysics can be uncovered in
unexpected places.


\section{Conclusion}

X-ray binaries emit across the electromagnetic spectrum, and a full
understanding of accretion processes in this environment depends on
multiwavelength observations. Not only do different wavelengths
illuminate different aspects of the behavior (inner disk, outer disk,
companion star, jet, etc), but also connections between the different
wavelengths provide essential information on causal connections
between them. Understanding X-ray binaries to the fullest extent
possible requires knowledge of not only X-ray and gamma-ray astronomy,
but also optical, ultraviolet, infrared, and radio wavelengths.


\section{Acknowledgements}

I am grateful to the Instituto de Astrof\'{i}sica de Canarias for the
invitation and funding to present a series of lectures at the 21st IAC
Winter School, on which this work is based. I would also like to
gratefully acknowledge Valerie Mikles, Chris Britt, Lauren Gossen, and
Chris Dupuis for providing an abundance of helpful comments on this
manuscript and catching many mistakes in earlier versions. This is the
document I would like to have given them when they began working in
this field. Preparation of this work has made extensive use of NASA's
Astrophysics Data System.

\section{Glossary of Objects Cited}
\label{ObjectSection}

\noindent
{\bf 2A~1822--371:} Persistent eclipsing neutron star LMXB.  
ADC source.

\noindent {\bf 2S~0921--630:} Persistent eclipsing neutron star LMXB.
ADC source. Long period.

\noindent
{\bf 4U~0614+091:} Persistent neutron star UCXB.

\noindent
{\bf 4U~1626--67:} Persistent neutron star UCXB. 

\noindent
{\bf 4U~1735-44:} Persistent neutron star LMXB.

\noindent {\bf 4U~1915--05:} Persistent neutron star UCXB. Prototypical dipping source.

\noindent {\bf 4U~2129+47:} Quasi-persistent eclipsing neutron star
LMXB. ADC source.  

\noindent
{\bf A~0620--00:} Prototypical transient black hole LMXB.

\noindent {\bf Cir~X-1:} Persistent neutron star IMXB or HMXB.
Eccentric orbit. Microquasar.

\noindent
{\bf Cyg~X-2:} Persistent neutron star LMXB.  Long period

\noindent {\bf EXO~0748--676:} Quasi-persistent eclipsing neutron star
LMXB. 

\noindent
{\bf GRO~J0422+32:} Transient black hole LMXB.

\noindent
{\bf GRO~J1655--40:} Recurrent transient black hole IMXB.  Microquasar.

\noindent {\bf GRS~1915+105:} Quasi-persistent black hole LMXB. Microquasar.  Very long period.

\noindent
{\bf GS~2000+250:} Transient black hole LMXB.

\noindent
{\bf GX\,339--4:} Recurrent transient black hole LMXB.  

\noindent
{\bf Her~X-1:} Persistent eclipsing neutron star IMXB.  Prototypical
warped disk source.

\noindent
{\bf LMC~X-4:} Persistent neutron star HMXB.

\noindent
{\bf M33~X-7:} Extragalactic eclipsing black hole HMXB.

\noindent {\bf Sco X-1: } Persistent prototypical neutron star LMXB. First
extra-solar X-ray source discovered.  Brightest persistent
extra-solar X-ray source in the sky.

\noindent
{\bf Ser~X-1:} Persistent neutron star LMXB.

\noindent
{\bf SMC~X-1:} Persistent neutron star HMXB.

\noindent {\bf Swift~J1753.5--0127:} Transient or quasi-persistent
black hole LMXB. 

\noindent {\bf X-ray Nova Muscae 1991:} Transient black hole LMXB.

\noindent {\bf V404~Cygni:} Transient black hole LMXB. Long period.

\noindent {\bf XTE~J1118+480:} Recurrent transient black hole LMXB.

\noindent
{\bf XTE~J1859+226:} Transient black hole candidate LMXB.  

\noindent
{\bf XTE~J2123--058:} Transient eclipsing neutron star LMXB.  

\section{Glossary of Acronyms}
\label{AcronymSection}

\noindent
\begin{tabular}{ll}
ACF & Auto-correlation function\\
ADC & Accretion disk corona\\
AGN & Active galactic nucleus\\
BATSE & Burst And Transient Source Experiment (on board CGRO)\\
BHXRT & Black hole X-ray transient\\
CCF & Cross-correlation function\\
CGRO & Compton Gamma-ray Observatory\\
CV & Cataclysmic variable\\
DIM & Disk instability model \\
DN & Dwarf nova\\
FRED & Fast rise exponential decay\\
HMXB & High-mass X-ray binary\\
HST & Hubble Space Telescope\\
IMXB & Intermediate-mass X-ray binary\\
IR & Infrared\\
IUE & International Ultraviolet Explorer\\
LMXB & Low-mass X-ray binary\\
\end{tabular}

\noindent
\begin{tabular}{ll} 
QPO & Quasi-periodic oscillation\\
RXTE & Rossi X-ray Timing Explorer\\
SED & Spectral Energy Distribution\\
STIS & Space Telescope Imaging Spectrograph (on board HST)\\
SXT & Soft X-ray transient\\
UCXB & Ultracompact X-ray binary\\
UV & Ultraviolet\\
\end{tabular}

\bibliography{hynes}
\bibliographystyle{cambridgeauthordate}

\end{document}